\documentclass[aip]{revtex4-1}
\usepackage{amsmath,amsfonts,amssymb,amscd,amsthm,amsbsy, color}
\usepackage{graphicx,epstopdf}
\usepackage{appendix}
\usepackage{wasysym}

\draft

\newcommand\beq[1]{ \begin{equation}\label{#1} }
\newcommand{\eeq}{ \end{equation} }
\newcommand{\beqno}{ \[ }
\newcommand{\eeqno}{ \] }
\newcommand\beqa[1]{ \begin{eqnarray} \label{#1}}

\newcommand{\eeqa}{ \end{eqnarray} }
\newcommand{\beqano}{ \begin{eqnarray*} }
\newcommand{\eeqano}{ \end{eqnarray*} }
\newcommand\equ[1]{{\rm (\ref{#1})}}

\newcommand{\eqi}[1]{$#1$}
\DeclareRobustCommand*{\unit}[1]{\def~{\,}\ensuremath{\mathrm{\,#1}}}

\begin{document}

\title{Orbital stability of ensembles of particles in regions of magnetic reconnection in Earth's magneto-tail}

\author{Christoph Lhotka}
\email{christoph.lhotka@oeaw.ac.at}
\affiliation{Space Research Institute, Austrian Academy of Sciences, Schmiedlstrasse 6,
A-8042 Graz}

\author{Philippe Bourdin}
\email{philippe.bourdin@oeaw.ac.at}
\affiliation{Space Research Institute, Austrian Academy of Sciences, Schmiedlstrasse 6,
A-8042 Graz}

\author{Elke Pilat-Lohinger}
\email{elke.pilat-lohinger@univie.ac.at}
\affiliation{Institute of Astrophysics, University of Vienna, T\"urkenschanzstrasse 17,
A-1180 Wien}

\date{today}

\begin{abstract}
We investigate the collective behaviour of particle orbits in the vicinity of
magnetic reconnection in Earth's magneto-tail. Various regions of different
kinds of orbital stability of particle motions are found. We locate regimes of
temporary capture of particle orbits in configuration space as well as
locations, where strong particle accelerations take place.  With this study we
are able to provide a detailed map, i.e. the topology, of high and low
acceleration centers close to the reconnection site. Quasi-regular and chaotic
kinds of motions of elementary particles can be determined as well.
The orbital stability of particle orbits is obtained by a statistical analysis
of the outcome of the system of variational equations of particle orbits within
the framework of particle-in-cell simulations. Using the concept
of Lyapunov Characteristic Numbers to ensembles of particle orbits we
introduce Lyapunov Ensemble {\bf Averages} to describe the response of particle
orbits to local perturbations induced by the electro-magnetic field.

\end{abstract}

\keywords{{\bf Lyapunov Ensemble Averages}, Magnetic reconnection, Earth magneto-tail}

\pacs{}

\maketitle


\section{Introduction}

Magnetic reconnection is a term that describes the fundamental change of
connectivity within magnetic field topologies.  \citet{Giovanelli:1946}
associated observations of energetic outbursts on the Sun with reconnection and
several other phenomena in the Earth's magnetosphere are probably driven by
reconnection, like auroral sub-storms \citep{Russell+McPherron:1973}.  While we
know from Maxwell's equations that it is impossible to cut and reassemble
magnetic field lines, the process of reconnection requires us to think beyond
this paradigm and it is a riddle until today, what exactly happens in the
micro-physical regime of reconnection.  In a more macroscopic sense, established
descriptions of the effects of reconnection comprise a quasi-static dissipative
mechanism \citep{Sweet:1958,Parker:1957} and model with discontinuities
\citep{Sonnerup:1970}.  After magnetic field lines reconnect, stress in the
field may be released by the retraction of the newly connected field lines,
which accelerates plasma together with the field and forms a slow shock within
the outflow \citep[][]{Petschek:1964}. Fundamental reconnection physics may be studied 
with the help of numerical simulations and in-situ observations in space
\citep{Paschmann:2013,Karimabadi+al:2013,Treumann+Baumjohann:2013,Treumann+Baumjohann:2015}.
It is still difficult to reach a non-collisional regime in laboratory
plasmas \citep{Zweibel+Yamada:2009,Yamada+al:2010}.
 
When we approach the reconnection site to spatial distances of about the
electron gyroradius or the current sheet thickness, we find strongly
non-Maxwellian electron velocity distributions (eVDFs)
\citep{Hoshino+al:2001,Bourdin:2017}.  We also find that there is an enhanced
non-gyrotropic behavior of electrons near the stagnation point of the
reconnection outflow, as well as the magnetic separatrix, which is the border
between four topological distinct magnetic fields that meet in the reconnetion
center \citep[see blue field line in pannel\ a) and dark-gray color in panel\
d) of Fig.\,1 in][]{Bourdin:2017}.  One may explain these non-gyrotropic orbits
as meandering motions of electrons that cross the current sheet multiple times
\citep{Horiuchi+Sato:1994,Ng+al:2012}, acceleration through electric fields
\citep[][]{Bessho+al:2014}, and deflection from strongly curved magnetic fields near
the reconnection center \citep[][]{Zenitani+Nagai:2016} that is also observed for
ions \citep{Nagai+al:2015}. 

In the current study we aim to identify different kinds of behaviour of
ensembles of particle orbits by means of the analysis of the system of
variational equations of motion. We find regions close to the reconnection
center, which support 'quasi'-regular or irregular (chaotic) kinds of orbital
motions.  These regions are important to better understand the mixing behaviour
of particle orbits in phase space. Our results are strongly related to the
study of the form of velocity distribution functions. It is clear that the
structure of these distributions strongly depend on the kind of particle orbits
that may cross the regions in space, for which these distibutions are
calculated.  Numerical and analytical studies have already been used to
understand the motions of elementary particles from a dynamical systems point
of view. In \citet{buechner1989} the ratio between the minimum radius of
curvature of a magnetic field line and the maximum Larmor radius for given
particle energy has been used to distinguish between regular and chaotic
motion.  The authors find that for ratios much larger than unity an adiabatic
invariant of motion exists, while for ratios close to unity this adiabaticity
breaks, and resonance overlapping between the fundamental periods of bounce-
and gyromotion introduces deterministic chaos into the problem. While in
\citet{buechner1989} the authors studied curvature alone, the combined action
of field curvature and magnetic shear on the dynamics has been investigated in
\citet{buechner1991}.  A reduction of the reconnection problem to a two-degree of freedom
Hamiltonian system can be found in \citet{efthy2005}, but with an application to
reconnection in the solar atmosphere. The authors use a perturbative approach
and construct a Poincar\'{e} surface of section based on a simplified 
Hamiltonian model to describe various aspects of particle dynamics.
The case of interactions of particles with multiple reconnecting current sheets 
has been investigated in \citet{Anastasiadis2008}. In \citet{Gontikakis2006}
different kinds of orbits have been found in a 3D Harris-type reconnecting
sheet where chaotic orbits lead to an escape by stochastic accelerations, regular
orbits leading to escape along the field lines of the reconnecting magnetic
component, and mirror-type regular orbits that are trapped on invariant tori.
Analytical formulae that provide the kinetic energy gain of particle orbits are
derived and validated by means of numerical simulations.

Our apporach to identify the different kinds of orbital motions of particles in
the vicinity of magnetic reconnection is based on the concept of Lyapunov
Characteristic Exponents (LCE), see e.g.\ \cite{1980Mecc...15....9B,
1984JMTAS......101F} which provides a quantitative estimate for chaos in case
of exponential divergence of two initially nearby trajectories. As the LCE is
an asymptotic quantity, which cannot be determined exactly, it is more common
to use terms like LCI (Lyapunov Characteristic Indicator) or LCN (Lyapunov
Characteristic Number) instead of LCE.  About a decade later,
\cite{1993CeMDA..56..307F, 1993CeMDA..56..315L, 1996CeMDA..64....1C,
1997CeMDA..67...41F, 2007PhyD..231...30S, 2004CeMDA..90..127S} started to
develop short-time and fast methods deduced from the LCE to distinguish between
regular and chaotic motion. These methods are briefly discussed in
\cite{dvolhobook} and more in detail in \cite{skok2010, skokos2016chaos}. 

Classical chaos indicators are defined with respect to a single orbit that
obeys a unique initial condition. They are very useful tools to obtain
information about the vicinity of an orbit. Our approach differs from these
classical chaos indicators as follows: our aim is to understand the collective
behaviour of ensembles of particles in real space rather than the vicinity
of single particle trajectories in phase space. An ensemble of particles is
usually described in terms of a given velocity distribution function, which can
be used to define a mean (bulk) velocity of the ensemble within a given region.
In analogy we propose to generalize the concept of single particle chaos
indicators and introduce the mean indicator over different particle
trajectories within a given region in real space that we call Lyapunov Ensemble
{\bf Average} ({\bf LEA}). In practice we calculate the mean variation of the tangent
vectors for a set of initial conditions in phase space that is confined to a
given region in real space.  The resulting number ({\bf LEA}) will serve as an
indicator for the given region rather than for a unique orbit, and will provide
a qualitative description of the dynamics of particles trajectories that enter
this specific region.

For technical reasons we require the method to be fast:

\begin{itemize}

\item[i)] Magnetic reconnection accelerates charged particles, which
generates mean currents to change the electric and magnetic fields. Eventually,
reconnection will end when the lowest magnetic energy state is reached. Our main
interest lies in the dynamical picture of particle orbits during magnetic reconnection.
The tracing of particle orbits shall end before the fields have significantly
changed.

\item[ii)] The vector fields are only given within a finite simulation
box in configuration space, but numerical simulations of particle orbits show
that particles may exit the box already after very short time. Therefore it is
desirable to determine the indicator before the particle exits the simulation
box, where the vector fields are valid, i.e. to avoid the influence of
numerical errors that are more pronounced at the boundaries of the simulation 
boxes.

\end{itemize}

For these reasons we define a new indicator based on the finite-time
approximations of LCEs, which we call Lyapunov Ensemble
{\bf Average} ({\bf LEAs}).

The mathematical set-up  of the problem can be found in
Sec.~\ref{matmod}, with the definition of {\bf LEAs} in
Sec.~\ref{s:LEN}, and a description of the particle-in-cell (PIC) simulation
data that we use in Sec.~\ref{s:PIC}. The main simulation results are described
in Sec.~\ref{s:simu}, the conclusions and summary of our study can be found in
Sec.~\ref{s:sumu}.

\section{Mathematical set-up}
\label{matmod}

Let $q/m$ denote the charge-over-mass ratio of a particle, i.e.  either
electrons with $m=m_e$ and $q=-q_e$ or protons with $m=m_p$ and $q=+q_e$, where $q_e$ stands
for elementary charge. The force of interest $\vec F$, acting on a particle of
electric charge $q$ subject to an electric and magnetic field, is given by :
\beq{LF}
\vec F = q\left(\vec E+\vec v\times\vec B\right) \ .
\eeq

\noindent Here,  $\vec E=\vec E\left(\vec r,t\right)$ and 
$\vec B=\vec B\left(\vec r,t\right)$ are vector fields that
depend on the position of the particle $\vec r=\vec r(t)$ at given
time $t$, $\vec v=\vec v(t)$ is the velocity $\vec v = d\vec r/dt$,
and the acceleration of the particle is simply given by:
\beq{eom}
\frac{d^2\vec r}{dt^2} = 
\frac{q}{m}\left(
\vec E\left(\vec r, t\right) + 
\frac{d\vec r}{dt}\times\vec B\left(\vec r, t\right)
\right) \ .
\eeq

\noindent Let $x,y,z$ and $\dot x, \dot y, \dot z$ denote the components of
the position and velocity vectors with $\vec r=\left(x,y,z\right)$ and $\vec v=
\left(\dot x, \dot y, \dot z\right)$. To simplify the ongoing exposition we introduce the
following notation: let $\left(X_1,X_2,X_3\right)$ be the positions
$\left(x,y,z\right)$ and $\left(X_4,X_5,X_6\right)$ be the velocities
$\left(\dot x, \dot y, \dot z\right)$ and denote by $\vec X=\left(X_1,\dots,X_6
\right)$ the state vector spanned in the $6$ dimensional phase space, that
we denote by $P_{\vec X}$ in the further discussion. In this set-up \equ{eom} can be reduced to a $6$-dimensional system of
first order ODEs:
\beq{eom2}
\frac{d\vec X}{dt} = \vec f(\vec X, t) \ ,
\eeq 

\noindent with initial conditions $\vec X(0)=(X_1(0),...,X_6(0))$. Here, vector
function $\vec f=\vec f(\vec X)$, using the notation $\vec
F=\left(F_x,F_y,F_z\right)$, is given by $\equ{LF}$ as follows:
\beqno
\vec f = \left(X_4, X_5, X_6, F_x/m, F_y/m, F_z/m\right) \ .
\eeqno

\noindent The elements of the Jacobian matrix ${\mathbf J}={\mathbf J}(\vec X)$
of \equ{eom2} are given in terms of $\vec X$ by:
\beq{jac}
{\mathbf J}_{ij}=\frac{df_i}{dX_j} \ , \quad i,j=1,\dots,6 \ ,
\eeq

\noindent where we used the notation $\vec f=(f_1,\dots,f_6)$. Let $\vec Y_i$
denote the deviation vector with $i=1,2,\dots,6$, and the matrix ${\mathbf Y}=(\vec
Y_1,\dots,\vec Y_6)$ be spanned by $\vec Y_i$. We choose $Y_{ij}(0)=\delta_{ij}$, 
with $i,j=1,\dots,6$, where $\delta_{ij}$ denotes the
Kronecker delta function with the property $\delta_{ij}=1$ for $i=j$ and
$\delta_{ij}=0$ otherwise.  The evolution in time of ${\mathbf Y}={\mathbf Y}(t)$ is 
given by the linearized system of equations of motion:
\beq{lin}
\frac{d\mathbf Y}{dt} = {\mathbf J}(t){\mathbf Y} \ ,
\eeq

\noindent with ${\mathbf J}={\mathbf J}(t)$ since ${\mathbf J}={\mathbf J({\vec
X}(t)})$.  We notice that the determination of ${\mathbf Y}$ with respect to
time $t$ requires to solve \equ{lin} together with \equ{eom2} totalling 42
ordinary equations of motion of first order in case of \equ{LF}.

\subsection{Lyapunov Ensemble Averages}
\label{s:LEN}

\begin{figure}
\begin{center}
\includegraphics[width=0.55\linewidth]{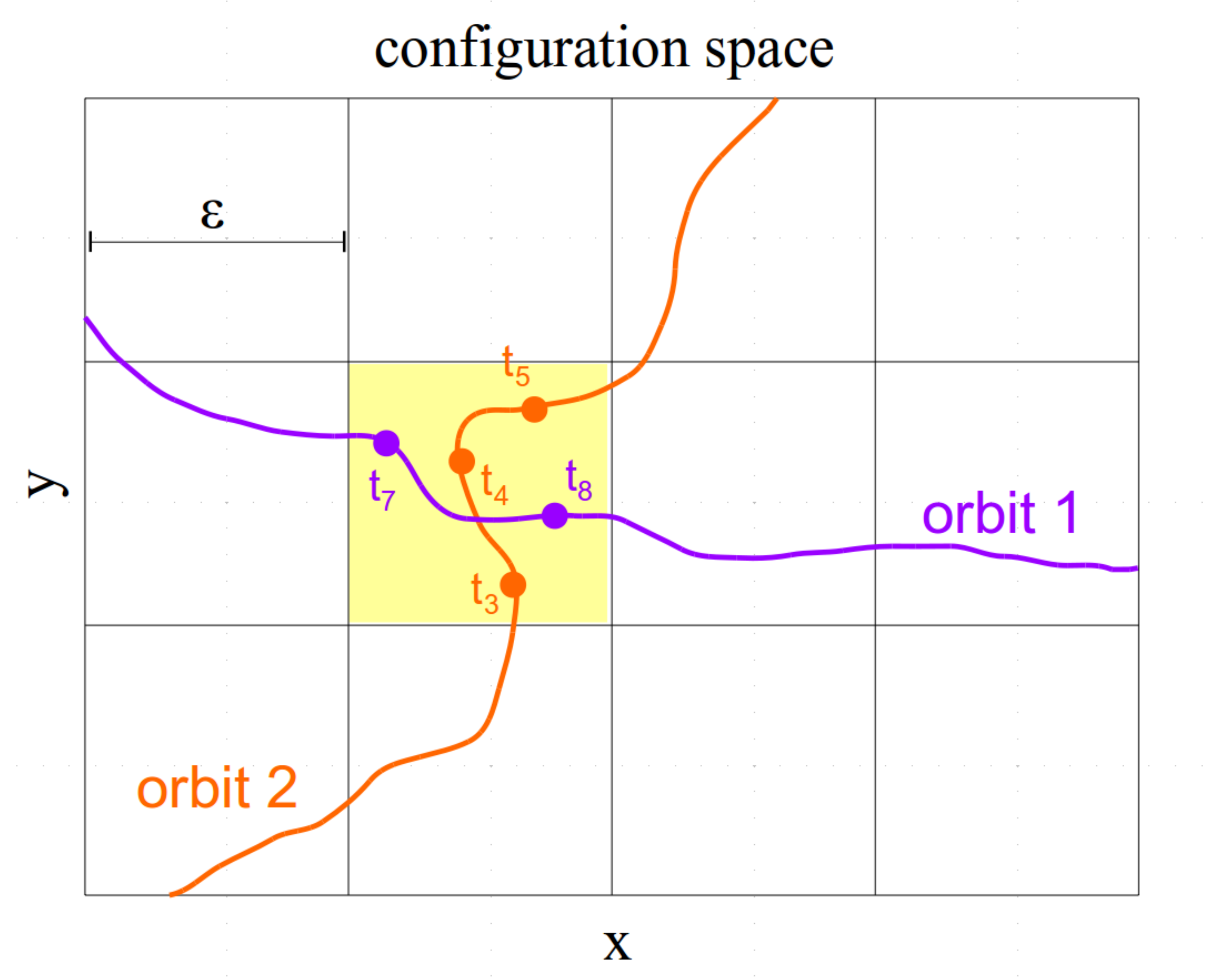}
\caption{Sketch of the method to calculate the {\bf LEA}. Two orbits cross the region 
$R_\varepsilon(\vec X_p)$ shown in yellow. Both orbits contribute with $\Lambda_n^{(m)}$ 
and $n=7,8$ for $m=1$ and $n=3,4,5$ for $m=2$ to the calculation of the {\bf LEA} within the 
yellow cell (see text for definitions of the symbols).}
\label{f:sketch}
\end{center}
\end{figure}

The development of the Lyapunov Ensemble {\bf Averages} ({\bf LEAs}) is motivated by the
work of \citet{1993CeMDA..56..307F, 1993CeMDA..56..315L}, where the authors
investigate the statistical moments of the time series of the local Lyapunov
Numbers. They show that the mean value of the distribution of these LLNs yield
the LCE of an orbit in case $N\to\infty$, where $N$ is the number of data
points of LLNs for the computed time evolution of a single orbit.  In
\citet{vogcon94} the authors investigate the distribution of the so-called
stretching numbers of single particle trajectories in dynamical systems. They
find that the spectrum of stretching numbers is independent of the
initial point along any particular orbit, that in a chaotic domain, it is
independent of the initial condition, and that the mean value over the spectrum
of stretching numbers tends to the LCE for N to infinity. In this study
the analysis is focused on single particle trajectories.  In our new
approach we perform the average over the LLNs of $M$ particle trajectories
crossing a given region instead. We provide a sketch of the method in
Fig.~\ref{f:sketch}. Two orbits (orbit 1, orbit 2) cross the region marked in
yellow at different time stamps $t_7$, $t_8$ (orbit 1) and $t_3$, $t_4$, $t_5$
(orbit 2). The {\bf LEA} for the yellow cell is then given by the average of the data
in the two time series that correspond to these times.  In practice thousands
of data points per cell are used in actual calculations.  The number is a mean
value of possible LLNs within a given region in real space and serves as an
indicator for the cell rather than for a single orbit itself.

A formal definition of the {\bf LEA} is as follows. Let $\Lambda=(\lambda_1,\dots\lambda_6)$ be the spectrum
of LCEs of the system \equ{eom2}, and $\Lambda_n$ be the LLNs obtained at time step $t_n$. We denote by
$(\Lambda_n)$, with $n=0,1,\dots,N$ the time series of sucessive approximations of $\Lambda$. In the 
limit $n\to\infty$ we have $\Lambda_n\to\Lambda$ which can be approximated by the mean over $(\Lambda_n)$ 
with $n=0,1,\dots,N$. To 
generalize the approach to multiple particle orbits we introduce the notation $(\Lambda_n)_m$ to be the
$m$-th time series of LLNs along the orbit of the $m$-th particle with $m=0,1,\dots, M$. We notice that at given time $t_p$ ($0\leq p\leq N$) 
the spectrum $\Lambda_p$ is linked to a position in phase space $\vec X_p\in P_{\vec X}$ since $\mathbf J$ 
in \equ{jac} also depends on $\vec X$. Let $R_\varepsilon(\vec X_p)\subset P_{\vec X}$ be the
$\varepsilon$-region centered around $\vec X_p$:
\beqno
R_\varepsilon(\vec X_p)=\{\vec X:\|\vec X-\vec X_p\|\leq\varepsilon\} \ .
\eeqno

\noindent Let $(\Lambda_n)_m=(\Lambda_0^{(m)},\Lambda_1^{(m)},\dots,\Lambda_N^{(m)}$),
and $\Lambda_n^{(m)}$ be the $n-th$ element of the $m-th$ time series of LLNs. We define the set ${\mathcal
M}^{\varepsilon}_p$ of ${\Lambda_n^{(m)}}$ passing through $R_\varepsilon(\vec
X_p)$ as follows:
\beqno
{\mathcal M}^{\varepsilon}_p = \{{\Lambda_n^{(m)}}:\vec X_n
\subset R_\varepsilon(\vec X_p)\} \ .
\eeqno

\noindent Within this context the spectrum of {\bf LEAs}
associated to a given point $\vec X_p$ is given by the mean values over all 
elements that belong to ${\mathcal M}^{\varepsilon}_p$: 
\beq{mlsn}
{\bar \Lambda_{p,\varepsilon}} = {\mu}\left({\mathcal M}^{\varepsilon}_p\right) \ .
\eeq

\noindent We notice that $\bar \Lambda_{p,\varepsilon}$ defined in terms of
\equ{mlsn} defines a spectrum in an  $\varepsilon$-region around $\vec X_p\in
P_{\vec X}$. In the limit $\varepsilon\to0$ the quantity $\bar \Lambda_{p,\varepsilon}$ reduces to the instant
value $\Lambda_p$ at given location $\vec X_p\in P_{\vec X}$ at time $t_p$
since $\vec X_p=\vec X(t_p)$ and by the existence and uniqueness theorem of
ordinary equations of motion. To ensure that enough data points cross
$R_\varepsilon(\vec X_p)$ small quantity $\varepsilon$ has to be choosen large enough. In the following, for
the sake of simplicity in the notation, we will denote by $\bar
\Lambda=(\bar\lambda_1,\dots,\bar\lambda_6)$ the {\bf LEA} for any point $\vec X_p\in
P_{\vec X}$ and for fixed value $\varepsilon=\varepsilon^*$ with
$\varepsilon^*>0$.

\subsection{Numerical Simulation setup}
\label{s:PIC}

The electric and magnetic fields in \equ{LF} are obtained from a PIC
simulation of anti-parallel magnetic fields using the open-source code
"iPic3D" \citep[][]{mark2010}.  We use the simulation data from
\cite{Bourdin:2017}, which provides a catalog of detailed electron velocity
distribution functions that are comparable with in-situ observations of
electrons in the tail of Earth's magnetosphere.

The PIC model is a two-dimensional (2D) setup that represents a reconnection
region in the magneto-tail of Earth, as well as certain magnetic field
configurations on the Sun.  The anti-parallel field lies within the plane of
the simulation, where the \eqi{x}-coordinate is parallel to the initial
background magnetic field \eqi{B_{\rm{0}} = 0.05477}.  We allow the particle
position vectors, particle velocity vectors, and the field vectors to have an
out-of-plane component, which is often called a 2.5D setup.  Our physical
domain spans over \eqi{25.6 \times 12.8\,d_{\rm{i}}^2} which we cover with
\eqi{512 \times 256} grid cells.  The ion inertial length \eqi{d_{\rm{i}}}
defines our characteristic length scale and is set by \eqi{B_{\rm{0}}} together
with the initial background number density \eqi{n_{\rm{0}} = 0.2}.  Our
computational grid distance is hence \eqi{\Delta\rm{r} = 0.05\,d_{\rm{i}}} and
we fill each 3D grid cell with \eqi{2^{15}} particles of each species,
electrons and protons.

We execute the simulation run on our in-house {\em LEO} computing cluster in
Graz, where we require a minimum of 8 compute nodes with 128 processors in
parallel and about \eqi{1.5\unit{TB}} of RAM in total for about 16\unit{hrs}.

The initial magnetic field and average particle velocities follow an analytical
Harris current sheet solution \citep{Harris:1962} that is supposed to remain
stable.  To trigger the reconnection exactly in the middle of the simulation
domain, we add a small perturbation to the initial fields.  Unlike earlier
works, our perturbation is much smaller in amplitude and spatial extent and
hence the evolution of the reconnection becomes more self-consistent
\citep{Bourdin:2017}.  We observe a free evolution of the reconnection until
about \eqi{t=22\,\Omega_{\rm{i}}^{-1}}, where \eqi{\Omega_{\rm{i}} =
{\rm{e\,B_{\rm{0}} / m_{\rm{p}}}}} is the ion gyrofrequency, \eqi{e} is the
charge of a proton, and \eqi{m_{\rm{p}}} is its mass.  Therefore, we take the
magnetic field data from a snapshot near \eqi{t_c=20.54\,\Omega_{\rm{i}}^{-1}},
when the reconnection is well developed but still in its free evolution phase.
The high-cadence magnetic and electric field data that we obtain from our PIC
simulation allows us to propagate particles in a realistic way with
time-interpolated field information.

For this study, we use ensembles of test particles that we propagate outside of
the main PIC simulation code, so that we are able to propagate also the
equation systems \equ{eom2} and \equ{lin}.  As our main PIC simulation run has
a very large number of particles, we do have an unprecedented low noise level
in our magnetic and electric fields, \eqi{\vec{B}} and \eqi{\vec{E}}.

\begin{figure}
\begin{center}
\includegraphics[width=0.45\linewidth]{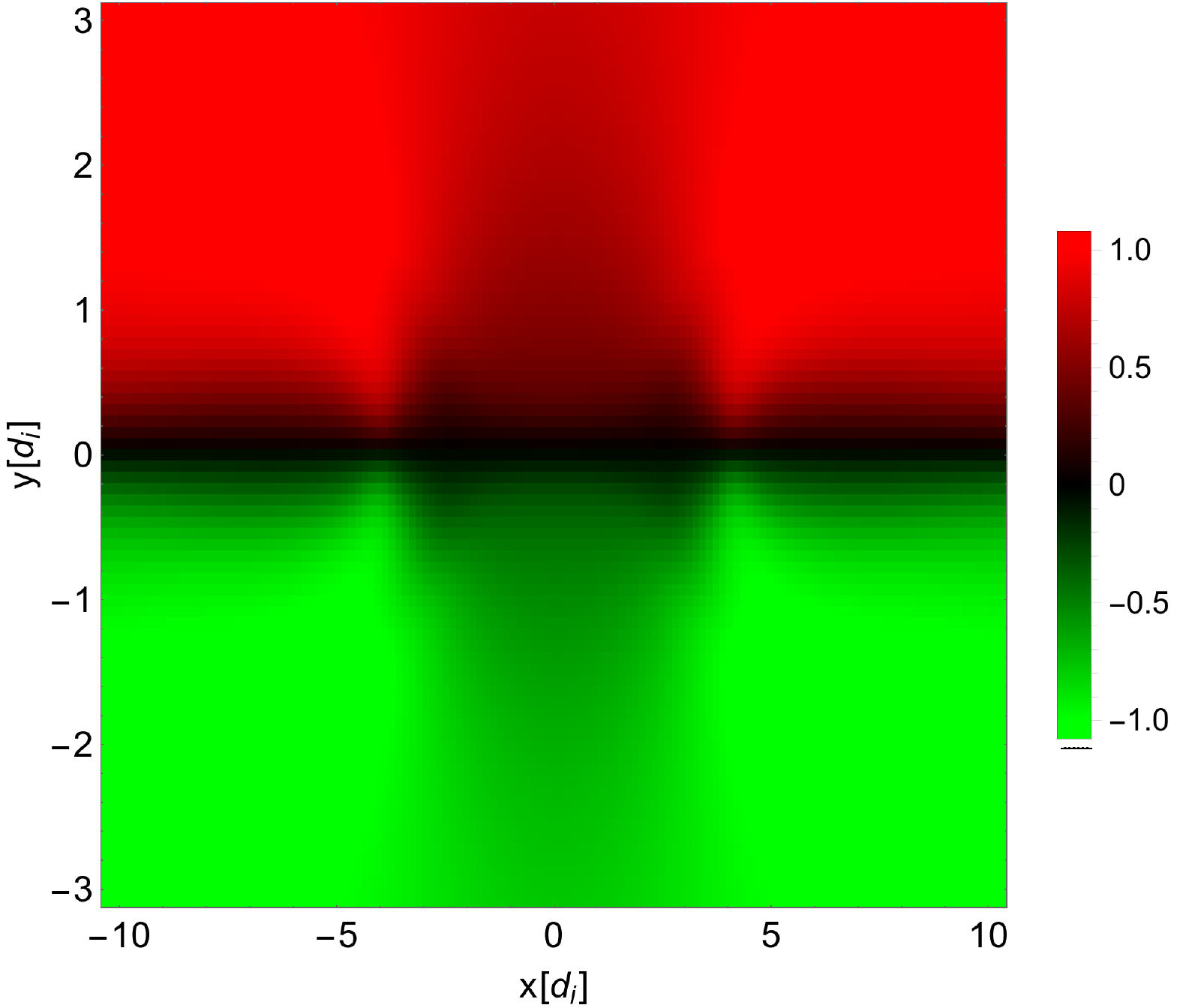}
\includegraphics[width=0.45\linewidth]{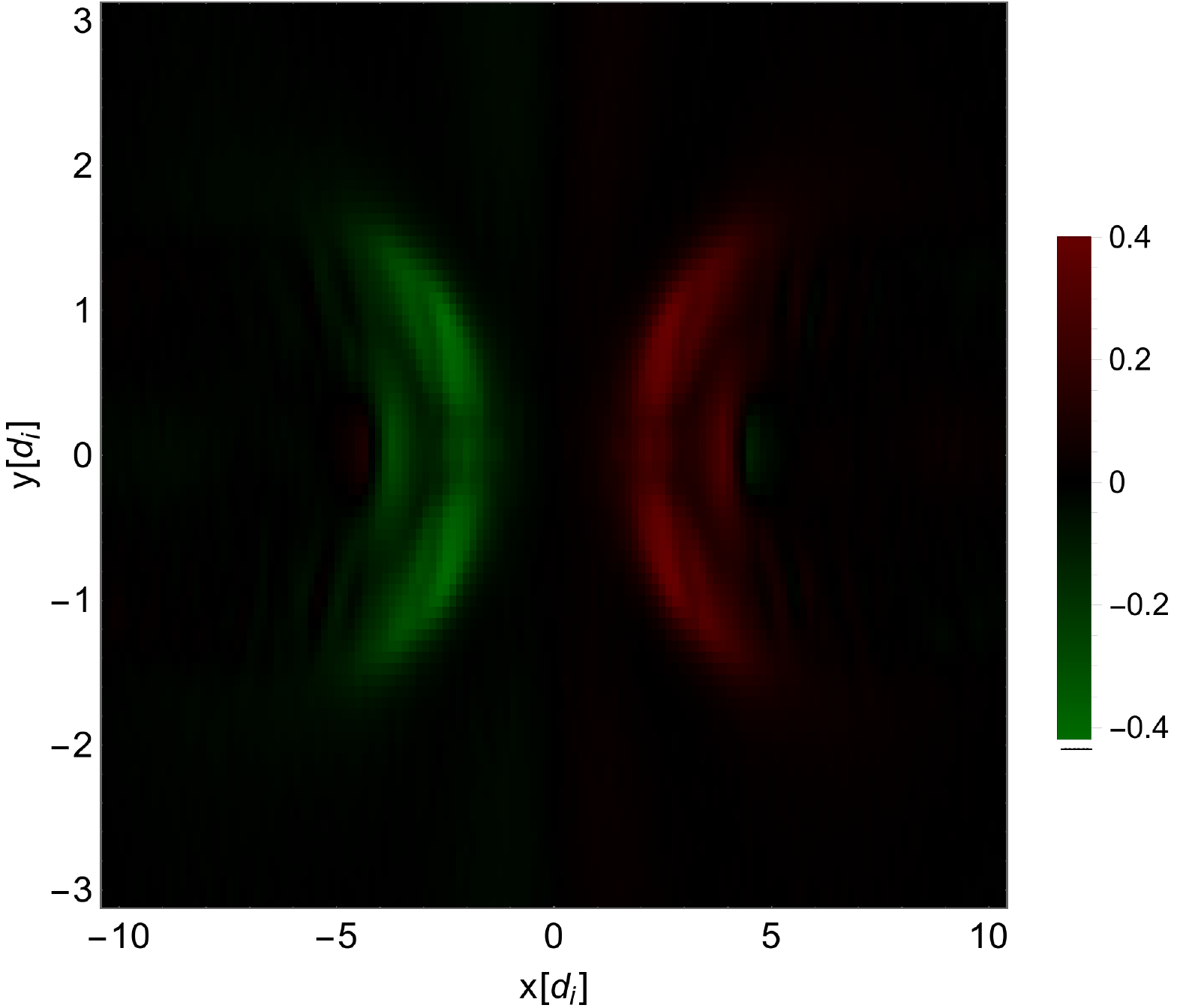}
\includegraphics[width=0.45\linewidth]{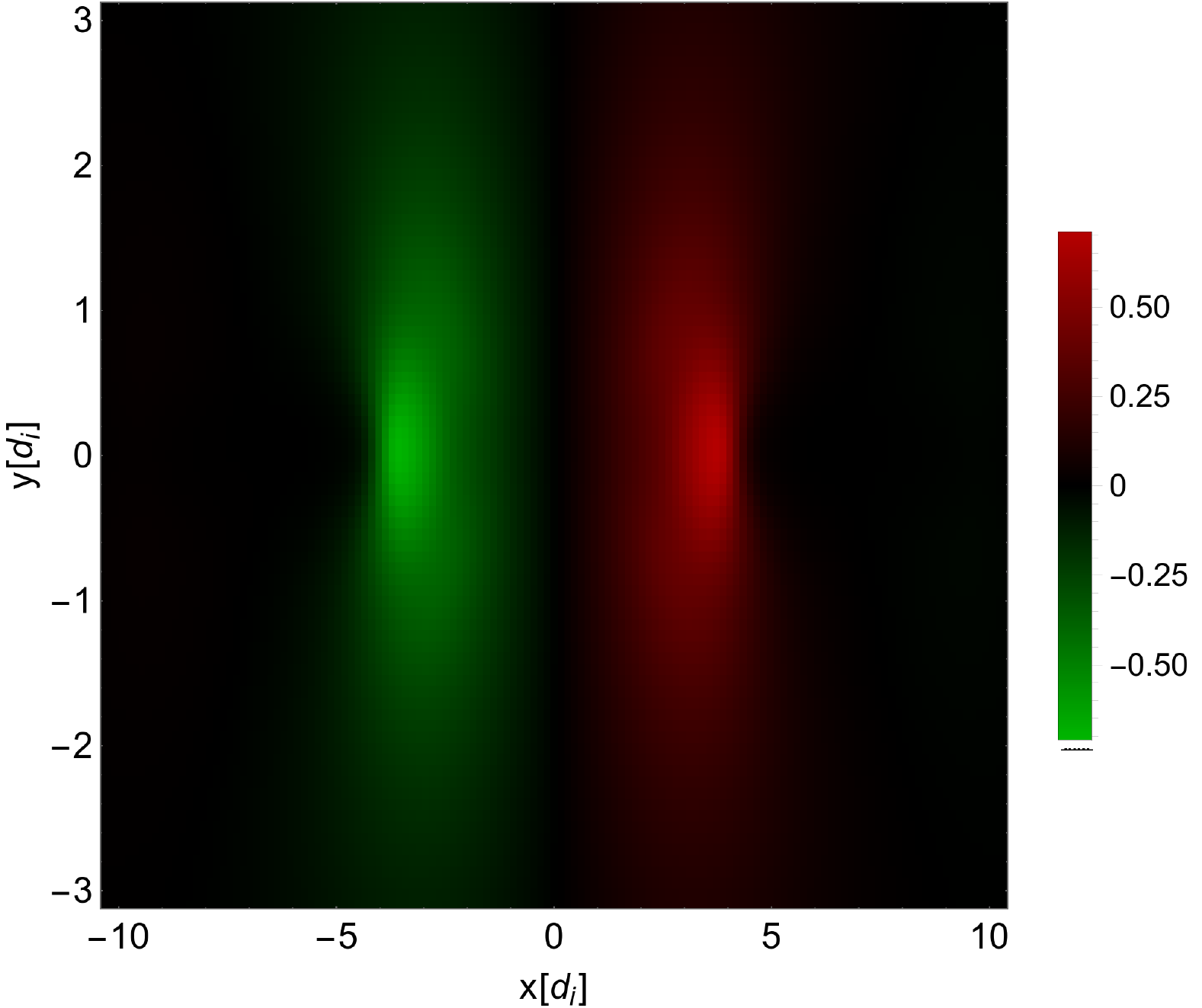}
\includegraphics[width=0.45\linewidth]{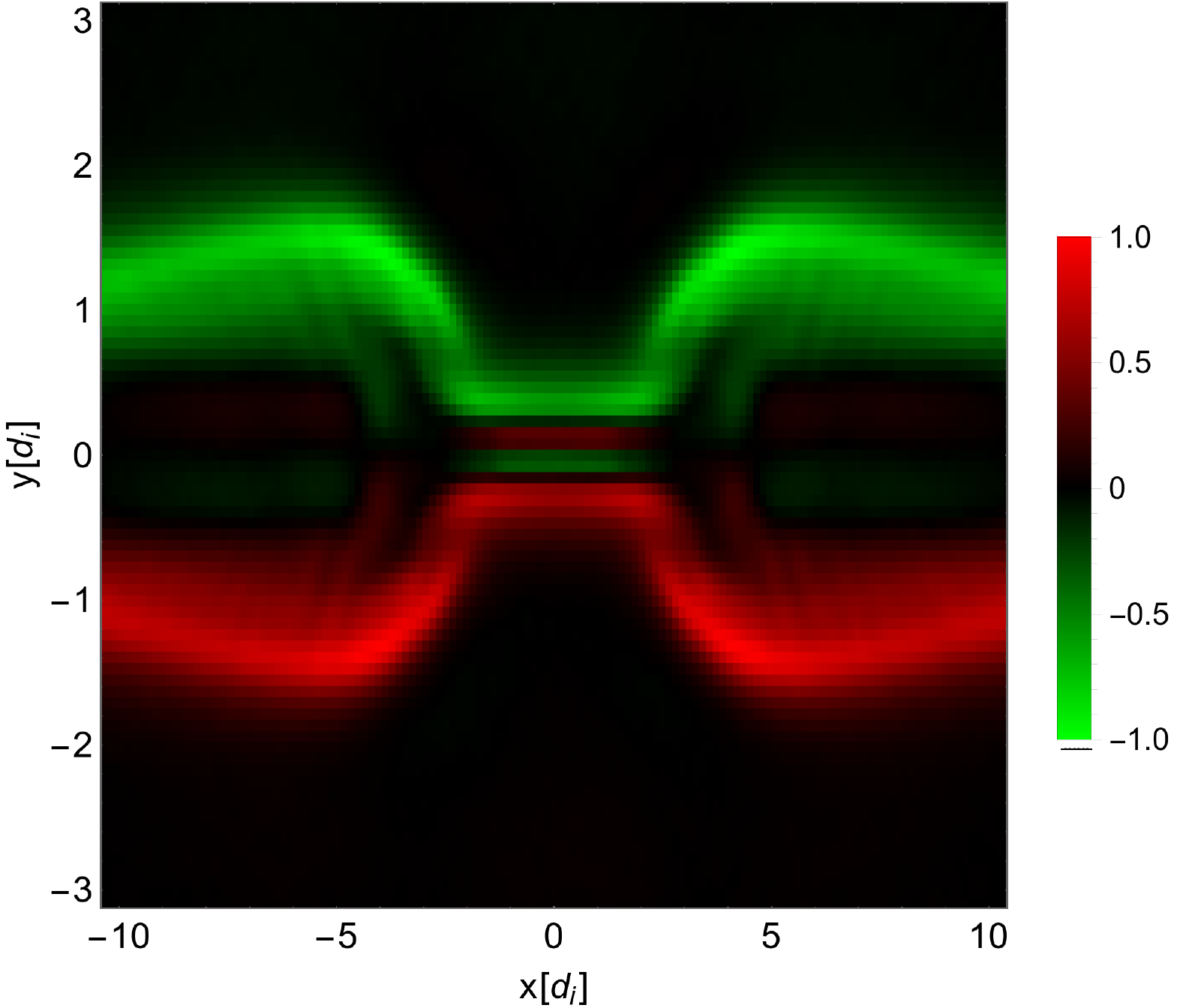}
\includegraphics[width=0.45\linewidth]{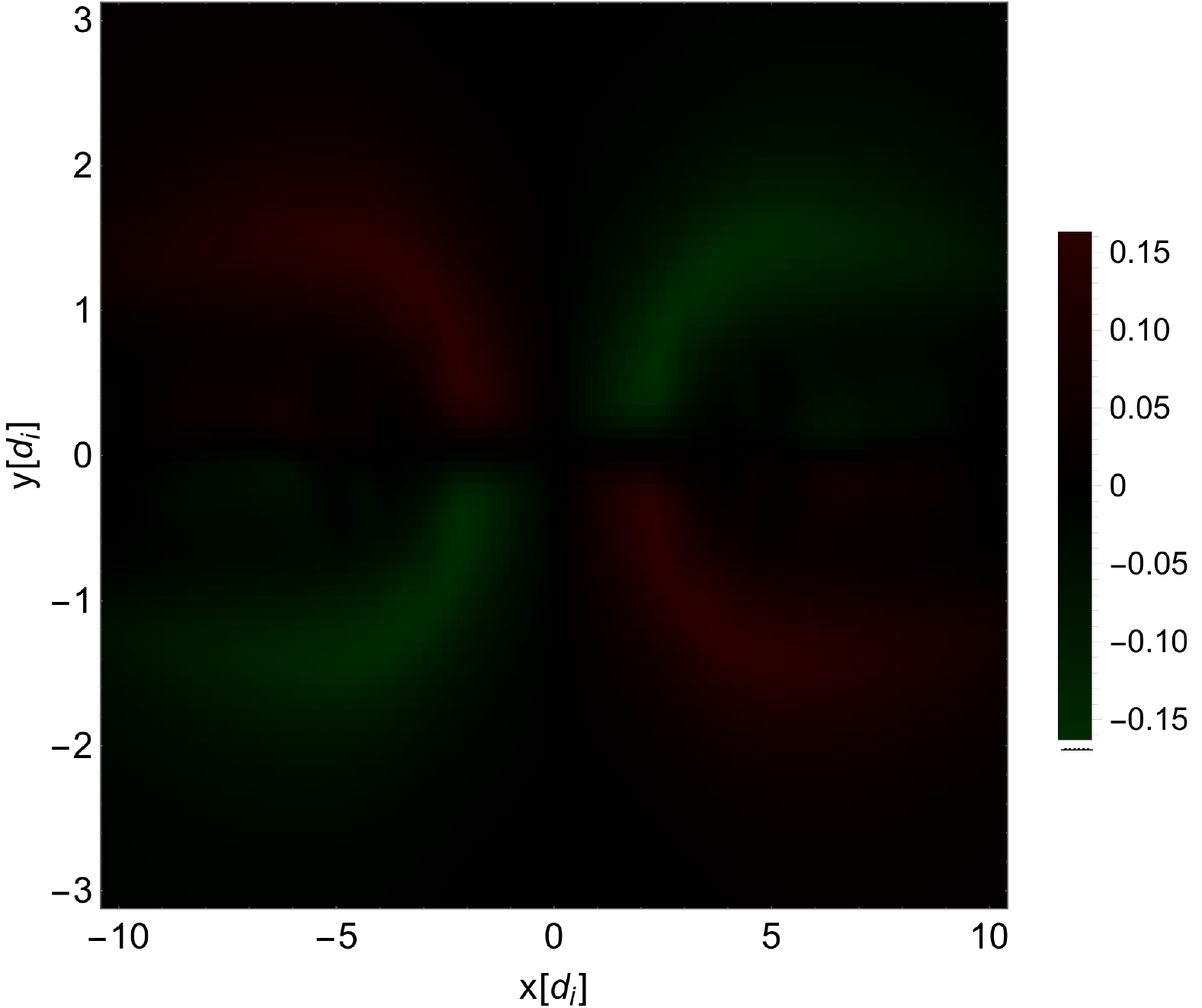}
\includegraphics[width=0.45\linewidth]{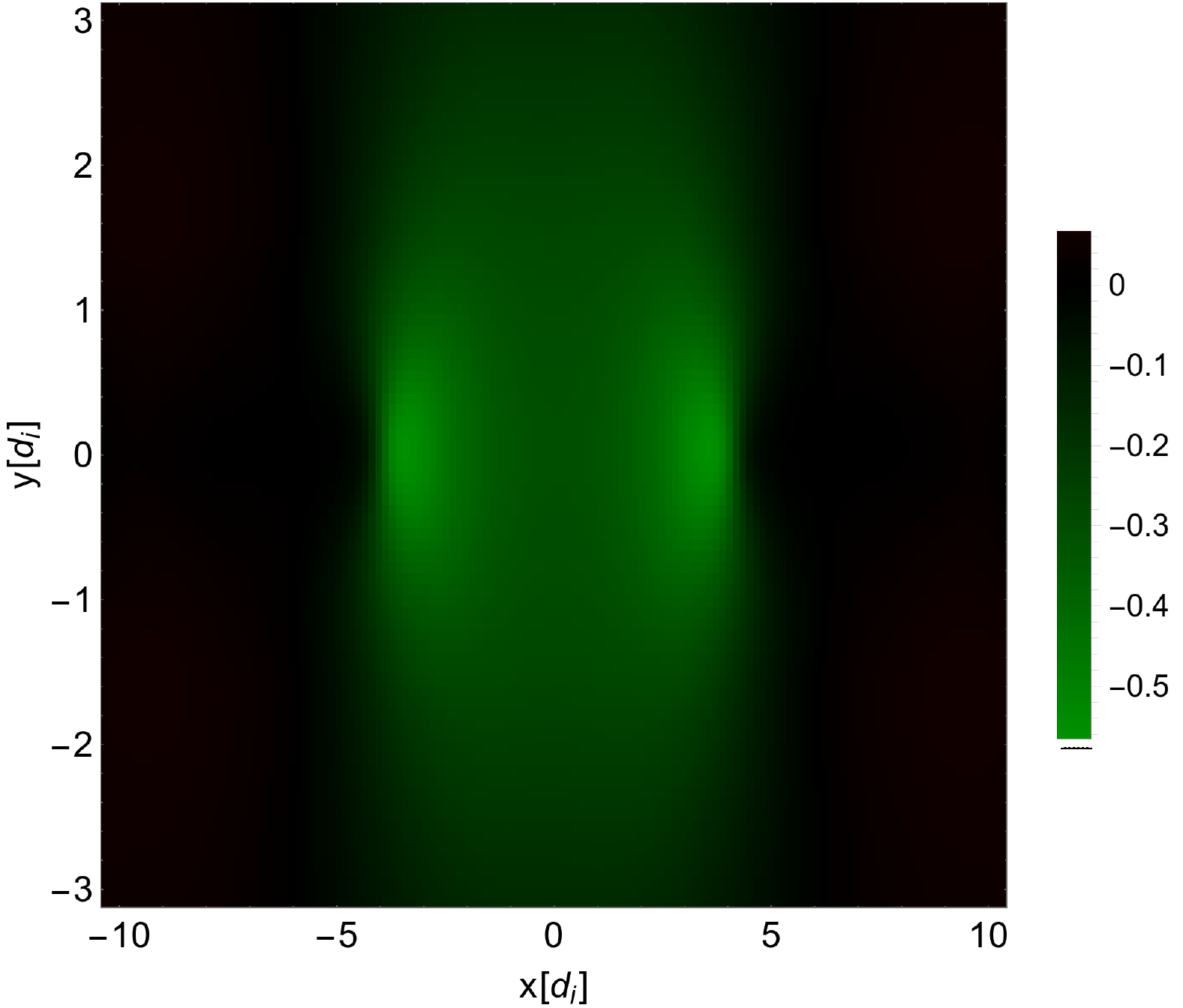}
\caption{Normalized components $x$ (top), $y$ (middle), and $z$ (bottom) of vector 
field $\vec B$ (left column) and $\vec E$ (right column), respectively.}
\label{f:fields}
\end{center}
\end{figure}

The vector fields $\vec E(\vec r,t)$ and $\vec B(\vec r,t)$ that enter \equ{LF}
are given from the PIC simulations on a discrete grid ($\vec r_i$, $t_i$ with
index $i$) only.  Let $I_{x}=(-L_x,L_x)$, $I_{y}=(-L_y,L_y)$, and
$I_{z}=(-L_z,L_z)$ denote three intervals with limits $L_x$, $L_y$, and $L_z$
(with $L_x=-12.8d_i$, $L_y=6.4d_i$, $L_z=0.5d_i$ in normalized units). The fields are
defined within $I=I_{x}\times I_{y}\times I_{z}$ on equally spaced
points with spacings $\Delta x$, $\Delta y$, and $\Delta z$ equal $0.05d_i$. We
use linear interpolation between the grid points on the space $I_{x}\times
I_{y}$ and constant periodic interpolation within the interval $I_{z}$ to
calculate $\vec E(\vec r,t)$ and $\vec B(\vec r,t)$ for any given point in $I$
with $|x|\leq L_x$, $|y|\leq L_y$. We notice that the interpolation routines
use divided differences to construct Lagrange or Hermite interpolating
polynomials. For the variational system \equ{lin} it also requires to evaluate
first order derivatives of the vector fields $\vec E$ and $\vec B$, i.e.
$d\vec E(\vec r,t)/dx$, $d\vec E(\vec r,t)/dy$, and $d\vec E(\vec r,t)/dz$, as
well as $d\vec B(\vec r,t)/dx$, $d\vec B(\vec r,t)/dy$, and $d\vec B(\vec
r,t)/dz$. The derivatives are calculated by piecewise symbolic differentiation
of the interpolating polynomials instead of the difference quotient rule.

\begin{figure}[!ht]
\begin{center}
\includegraphics[width=0.49\linewidth]{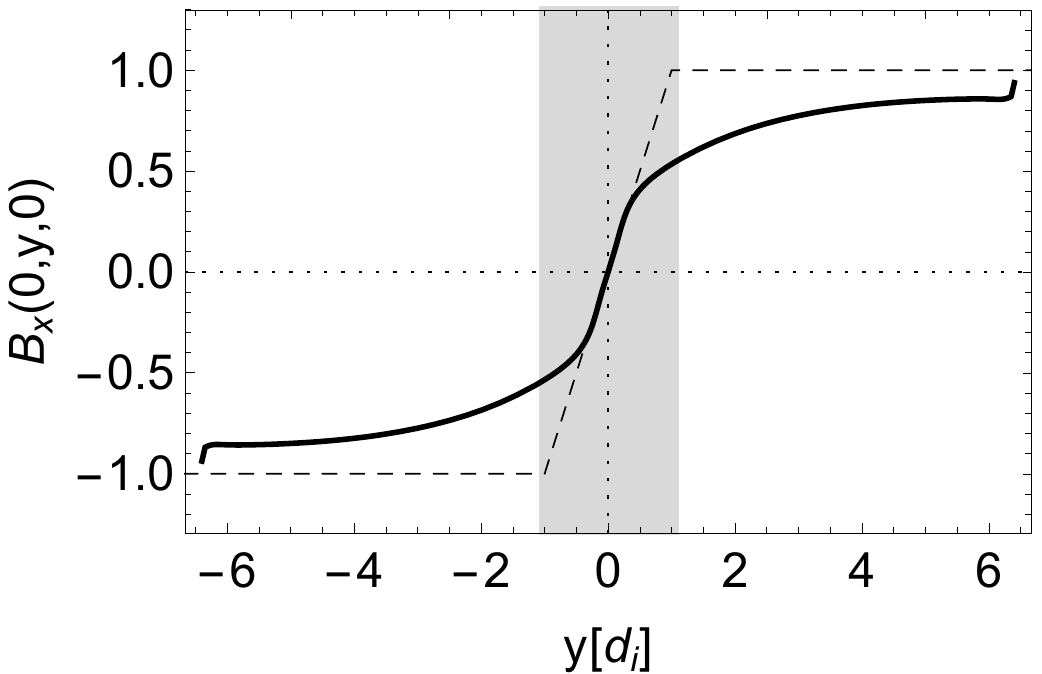}
\caption{Normalized $B_x(0,y,0)$ component of the magnetic field $\vec B$ versus $y$.
Dashed lines correspond to a Haris type model, see text.}
\label{f:Bfield}
\end{center}
\end{figure}

\noindent We fix the vector fields with respect to time $t$ in the following
discussion. The approximation is valid since the time that single particle orbits 
need to cross the simulation domain is much shorter than the timescales on 
which the electric and magnetic fields change.
For the sake of simplicity we use the notation $\vec E(x,y,z)=\vec
E(x,y,z,t_c)$ and $\vec B(x,y,z)=\vec B(x,y,z,t_c)$ from now on.  The
interpolated vector fields are shown in Fig.~\ref{f:fields}.  Fields have been
normalized by the background field strength $B_0$, and the maximum electric
field strength $E_0$.  A comparison of our vector field with a Haris type
reconnecting current sheet model \citep[see, e.g.][]{1993SoPh..146..127L} is
shown in Fig.~\ref{f:Bfield}. The plot shows the quantity $B_x(y)=B_x(0,y,0)$
in the interval $I_y$. The dashed line corresponds to a Haris type
approximation of the component $B_x(y)$. It perfectly agrees with the vector
field close to the reconnection center (gray-shaded region), while away from
the center the analytical model only qualitatively reproduces the results. We
remark, that in simplified models the electric vector field is usually
approximated by a constant $E_0$, i.e.  by $\vec E=(0,0,E_0)$, which is valid
in comparison with the component $E_z$ in Fig.~\ref{f:fields}. However, in the
other components, we clearly see that a sign change, e.g. in
$E_y(y)=E(x=0,y,z=0)$, has to be taken into account for changing sign in $y$.

The system of differential equations \equ{eom2} and \equ{lin} is solved in
Wolfram Mathematica (Version 11) using the NDSolve framework using 
ExplicitRungeKutta method with difference order 8 and variable step size 
control. The numerical
integrator has been adapted to incorporate the calculation of the {\bf LEAs} at given
time steps, $\Delta t=0.1$, of the numerical integration by direct access to the NDSolve
data structures. The initial conditions are taken
within the region $x(0)\in(0,10)d_i$ and $y(0)\in(0,5)d_i$ with $z(0)=0d_i$ and initial
velocities $\dot x(0)$, $\dot y(0)$ are obtained from a two-dimensional
Maxwellian distribution, while $\dot z(0)$ is taken to be zero for all initial
conditions. The initial deviation vectors $\vec Y_i$ for each initial condition
$\vec X(0)$ are taken to be $Y_{ij}(0)=\delta_{ij}$, $i,j=1,\dots,6$ (see
Sec.~\ref{matmod}).  After each time step $n\Delta t$, with $\Delta t=0.1$ an
orthogonal basis $\vec Y_i^*(n \Delta t)$ is obtained from $\vec Y_i(n \Delta
t)$ by means of a Gram-Schmidt orthogonalization process. The spectra
$(\Lambda_n)$ are then calculated from the norm of the projections of $\vec Y_i(n
\Delta t)$ onto $\vec Y_i^*(n\Delta t)$, which is also used as the new initial
condition to continue the integration of \equ{lin}. The resulting time series
$(\vec \Lambda_n)_m$ serve as the basis for the calculation of the {\bf LEAs} as
outlined in Sec~\ref{matmod}.  Integration is stopped either when the particle leaves
the rectangular region $|L_x|\times|L_y|$ or when
the integration time exceeds the limit where the approximations
of the vector fields $\vec E$, $\vec B$ are not valid anymore.

The accuracy of the methodology relies on the accuracy of the vector fields 
$\vec B$, $\vec E$ that have been obtained from PIC simulations with a very high
number of particles and a low noise in the electric fields. We do see fluctuations in the 
electric fields of the order of $10^{-5}$, that are due to the finite number 
of particles that have been used per grid cell. The total energy in our PIC
simulations is conserved with a precision of $0.18\%$. We also checked that the 
dominant component of the magnetic field $B_x(0,y,0)$ is in agreement with a Harris 
type model (see Fig.~\ref{f:Bfield}). To estimate the numerical error
of the vector fields and their partial derivatives with respect to $x$, $y$, $z$, we 
took the magnitude of the non-zero value of the divergence of the vector field 
$|\nabla\cdot\vec B|\simeq10^{-5}$ evaluated at the reconnection center $x=y=z=0$.
For the numerical integration of \equ{eom2} and \equ{lin} the precision of the integration
method was set to $10^{-10}$, which is still double the orders of magnitude than the calculated 
error of the vector field at the reconnection center. We notice that the design of the {\bf LEA} method does not require
long-term simulations of individual particle trajectories if one ensures that enough particles
are taken into account in the averaging process. The numerical errors in the integration
will therefore not grow too much as long as we set the maximum time of integration small 
enough, which is true in our case, since it is limited by the fact that the electrons exit the 
PIC simulation box already after short time.

\section{Numerical Simulations and Results}
\label{s:simu}

\begin{figure}[!ht]
\begin{center}
\includegraphics[width=.75\linewidth]{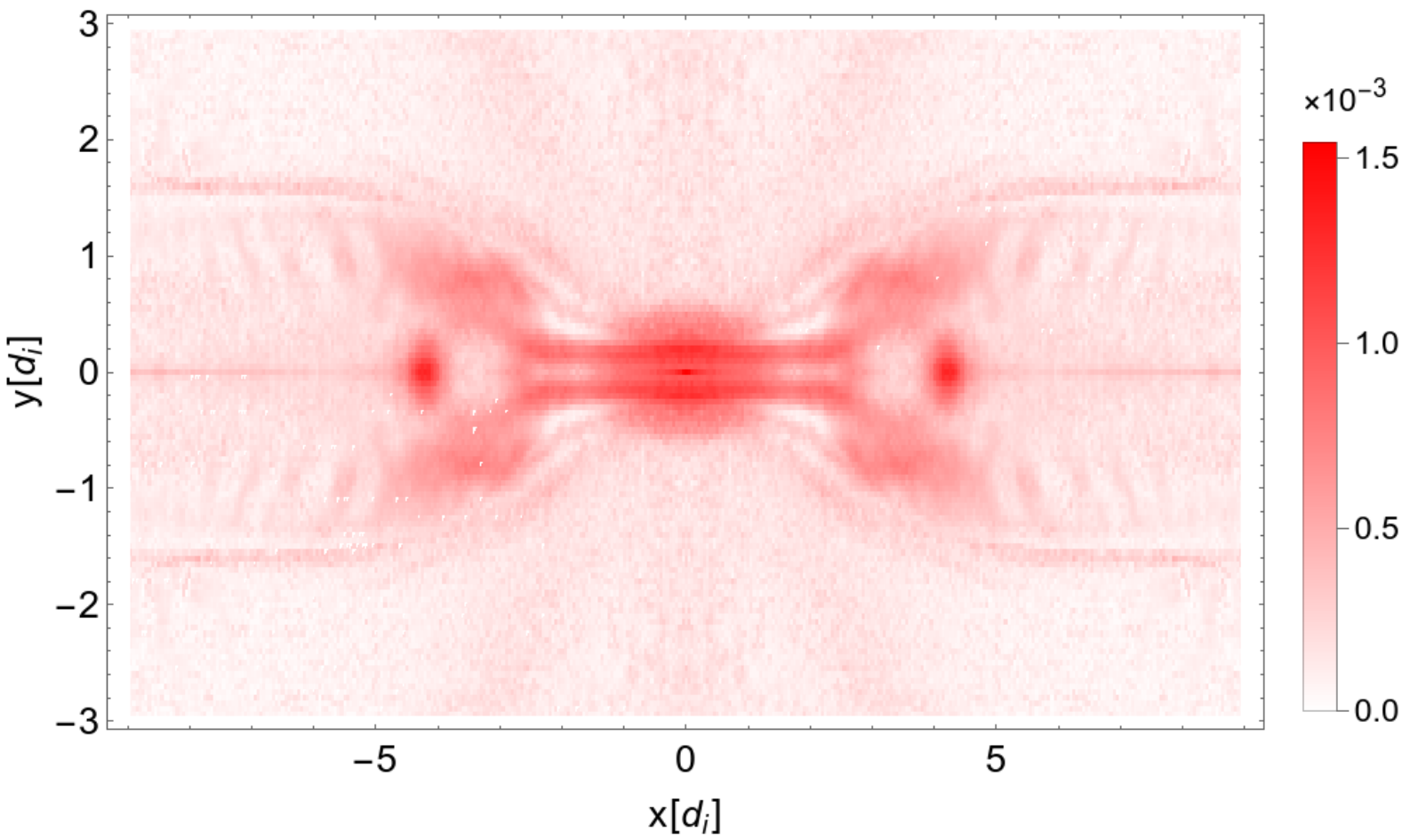}
\caption{Maximum {\bf LEA} number $\bar\Lambda_{max}$ projected to the space $I_x\times I_y$.}
\label{f:LEN}
\end{center}
\end{figure}

Our results are given in terms of the spectrum
$\bar\lambda_1,\dots,\bar\lambda_6$.  In the following we investigate different
quantities derived on the basis of this complete set of {\bf LEAs}. We first
investigate the quantity $\bar\Lambda_{max}=\max_{i=1,\dots,6}\bar\lambda_i$
which serves as an indicator for the overall stretching / contracting behaviour
of ensemble of particles. This number is shown in color code for different
cells with $\Delta x=\Delta y=0.05d_i$, and projected to the $(x,y)$-plane in
Fig.~\ref{f:LEN}. We clearly see that $\bar\Lambda_{max}$ is positive or zero
throughout $(I_x\times I_y)$.  This indicates that on the mean particles within
ensembles experience divergence and growth of local perturbations throughout
the reconnection center with high probability.  Still, the strength of growth
is not uniform. We find regions of variable magnitudes in $\bar\Lambda_{max}$
with peaks at various places, i.e. at the reconnection center or around the
island located at about $x=4.5d_i$, $y=0d_i$.  In other regimes
$\bar\Lambda_{max}$ is fluctuating, e.g. within the region inside the
separatrix, or the central inflow regime.  

\begin{figure}[!ht]
\begin{center}
\includegraphics[width=.45\linewidth]{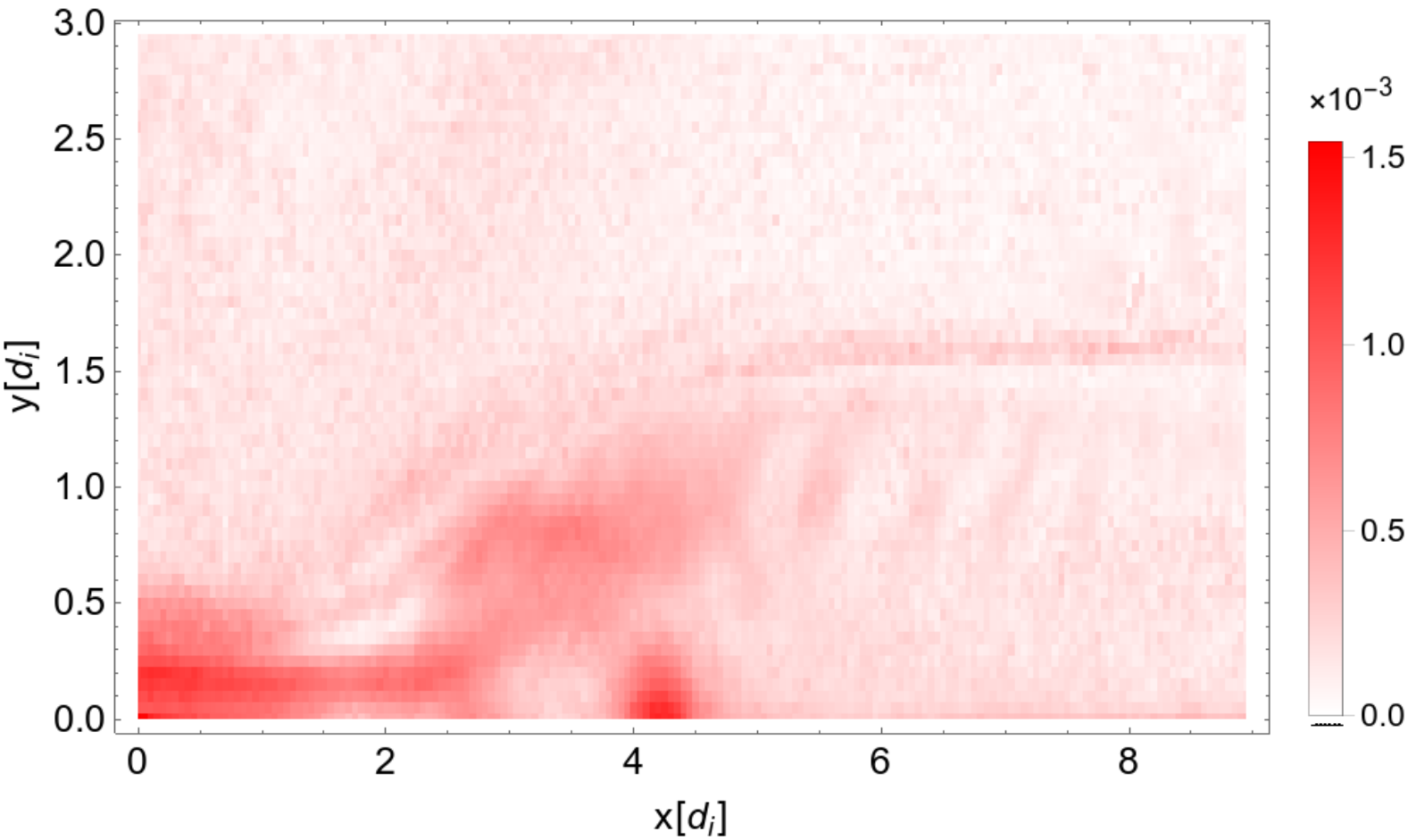}
\includegraphics[width=.45\linewidth]{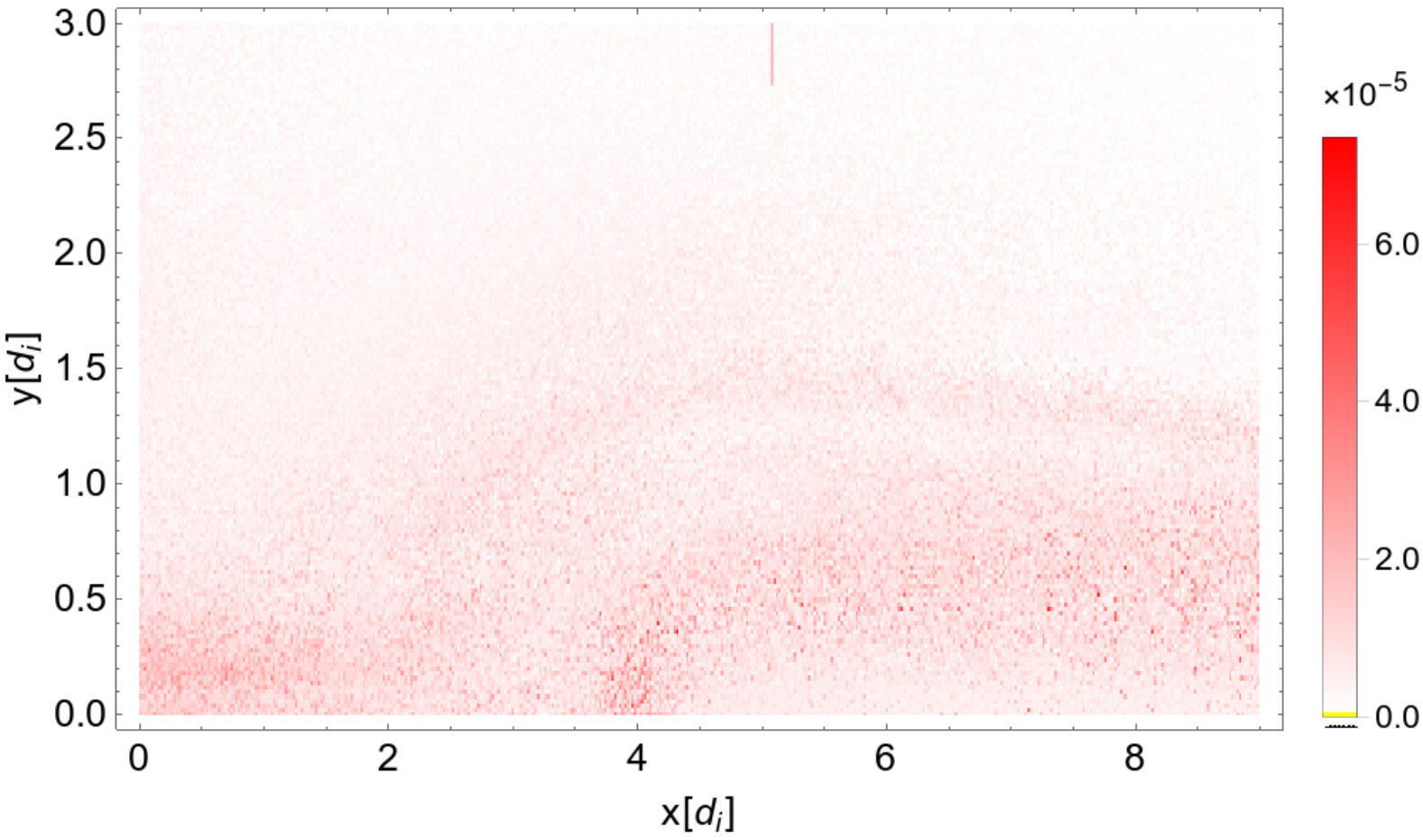}
\caption{Comparison of {\bf LEA} (left) with FLI (right).}
\label{f:FLI}
\end{center}
\end{figure}

{\bf Our first test in this work is to check the similarity of the LEA with the
original} Fast Lyapunov Indicator (FLI, \citet{1984JMTAS......101F}) in
Fig.~\ref{f:FLI}.  The left panel shows a magnification of the upper right part
of Fig.~\ref{f:LEN} and the right panel displays the FLI results based on the
same data set.  The comparison of both figures clearly shows that the most
pronounced features are visible in both plots, but are less visible in the FLI
figure, which can be explained by the fact that the FLI is calculated for a
single trajectory while the {\bf LEA} is a statistical quantity derived on the
basis of multiple orbits.  {\bf We notice that this test may be incomplete,
because the FLI is not based on local ensemble averages, like our LEA method
is. Therefore, future work should use several chaos indicators\cite{skok2010,
skokos2016chaos}, like GALI, FLI, MEGNO, RLI, and SALI, averaged locally in the
same way as our LEA approach.} Fig.~\ref{f:LEN} should also be compared with
the panels of Fig.~2 of \citet{Bourdin:2017} (we remark, that the coordinates
$y$ and $z$ need to be exchanged for the comparison). First, we clearly observe
that the locations of the main features of Fig.~\ref{f:LEN} are also present in
\citet{Bourdin:2017}.  Most visible the reconnection electric field (panel c))
is strongest in magnitudes at locations, where the dark red regions appear in
Fig.~\ref{f:LEN}.  We conclude, that the effect of the perturbations due to the
reconnection electric field is most efficient on enesembles of particle orbits,
where the {\bf LEAs} peaks in Fig.~\ref{f:LEN}.  The consequence of these
findings on the dynamics of ensembles of particle orbits is as follows: in dark
red regimes strong accelerations take place and lead to an increase of local
perturbations in magnitudes, which then results in changes in the evolution of
individual orbits (see also show cases at the end of this section). On the
contrary, perturbations in light red to white regions of the space $(x,y)$ are
less effective.

\begin{figure}[!ht]
\begin{center}
\includegraphics[width=.75\linewidth]{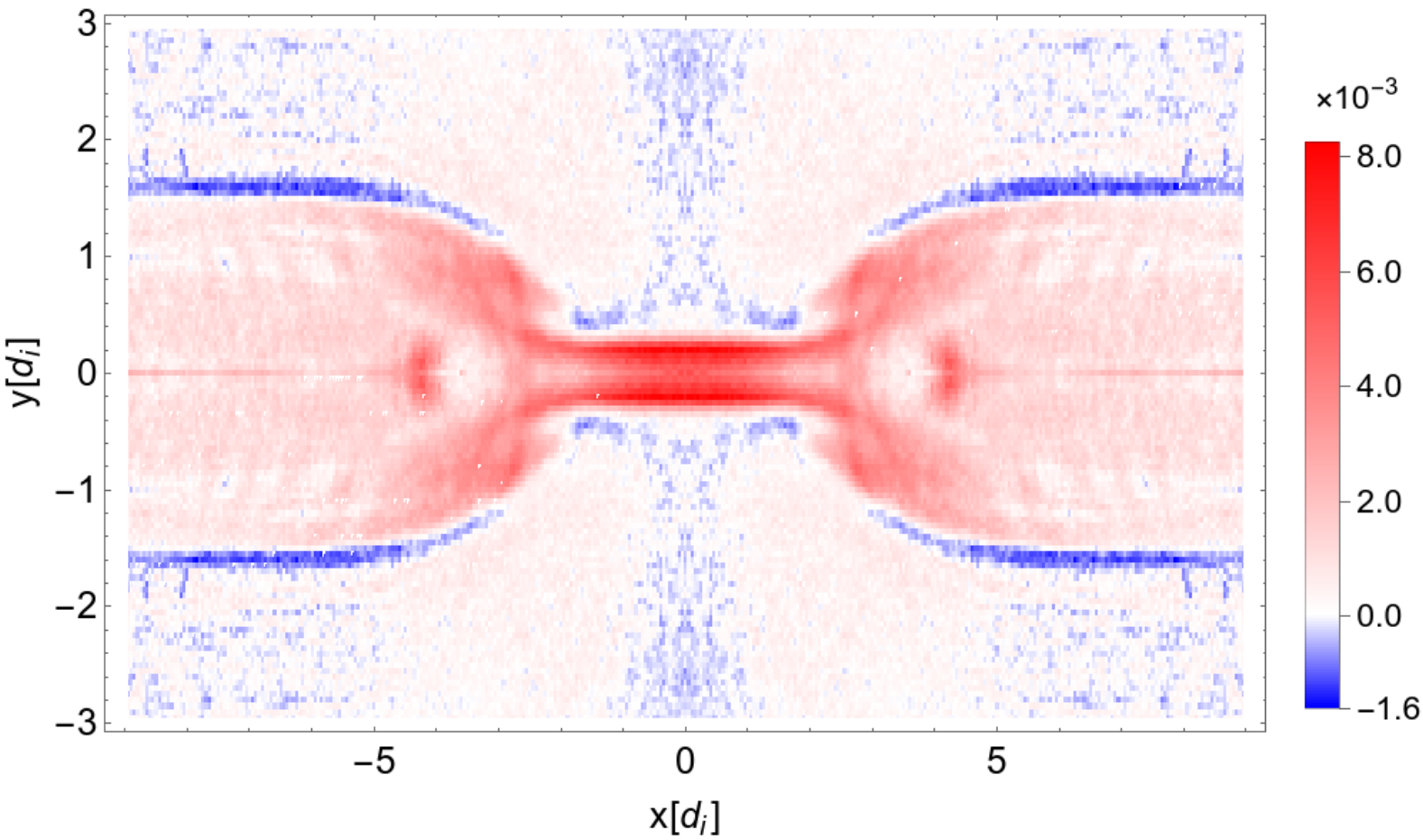} \\
\includegraphics[width=.75\linewidth]{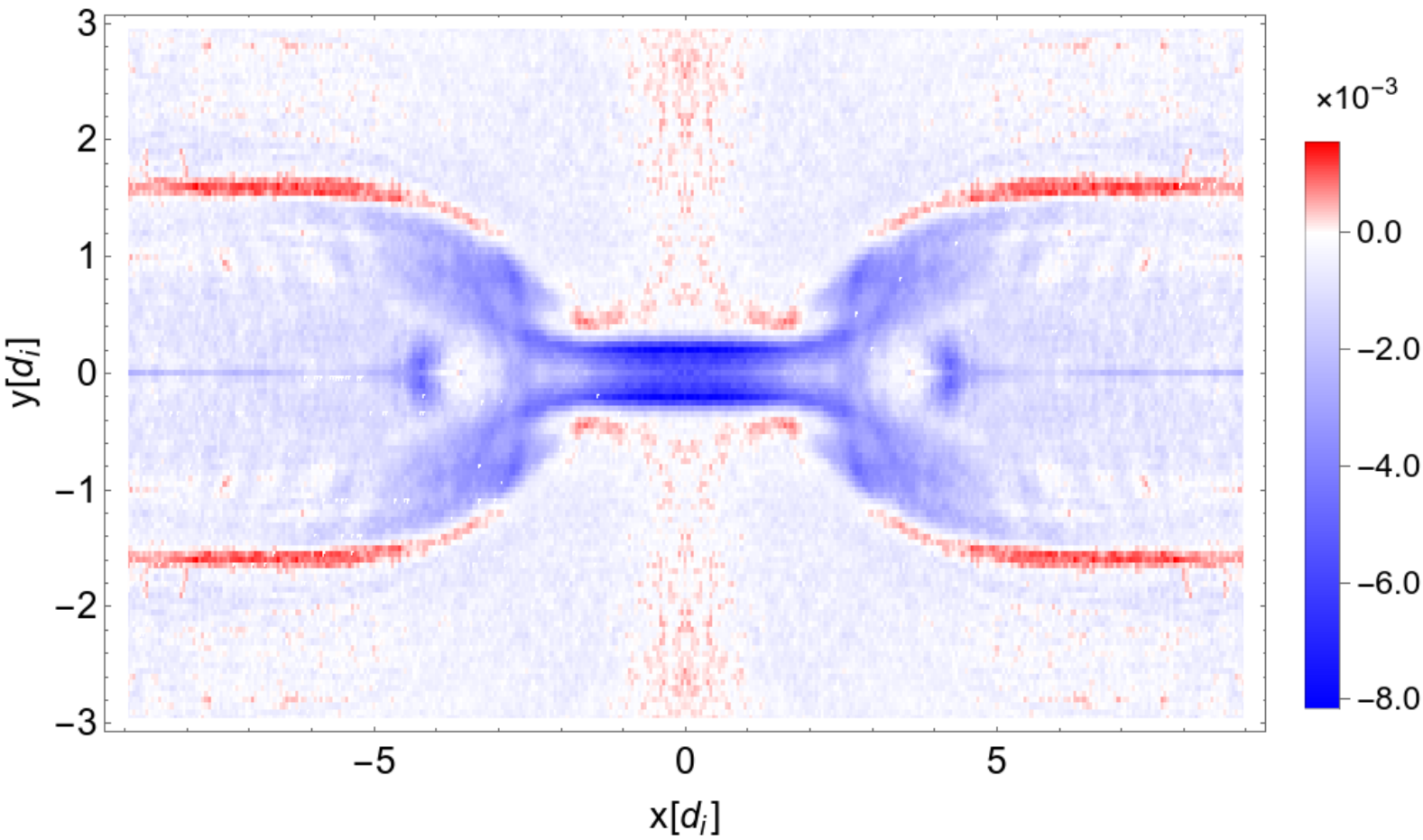}
\caption{Comparison of the quantity $\bar\Lambda_{123}$ (top) and the quantity
$\bar\Lambda_{456}$ (bottom) projected to the space $I_x\times I_y$.}
\label{f:LEN-LENv}
\end{center}
\end{figure}

\begin{figure}[!ht]
\begin{center}
\includegraphics[width=.48\linewidth]{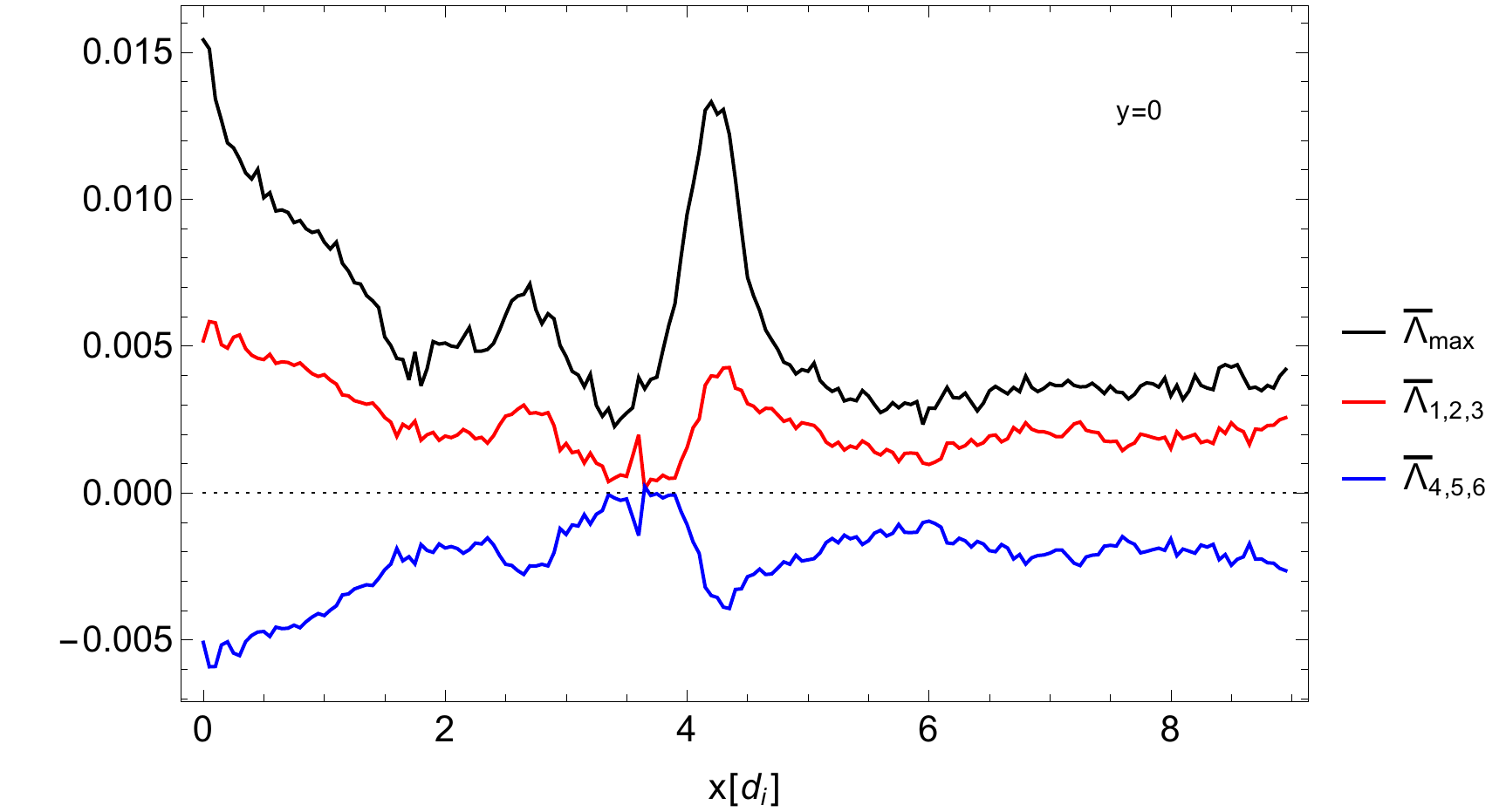}
\includegraphics[width=.48\linewidth]{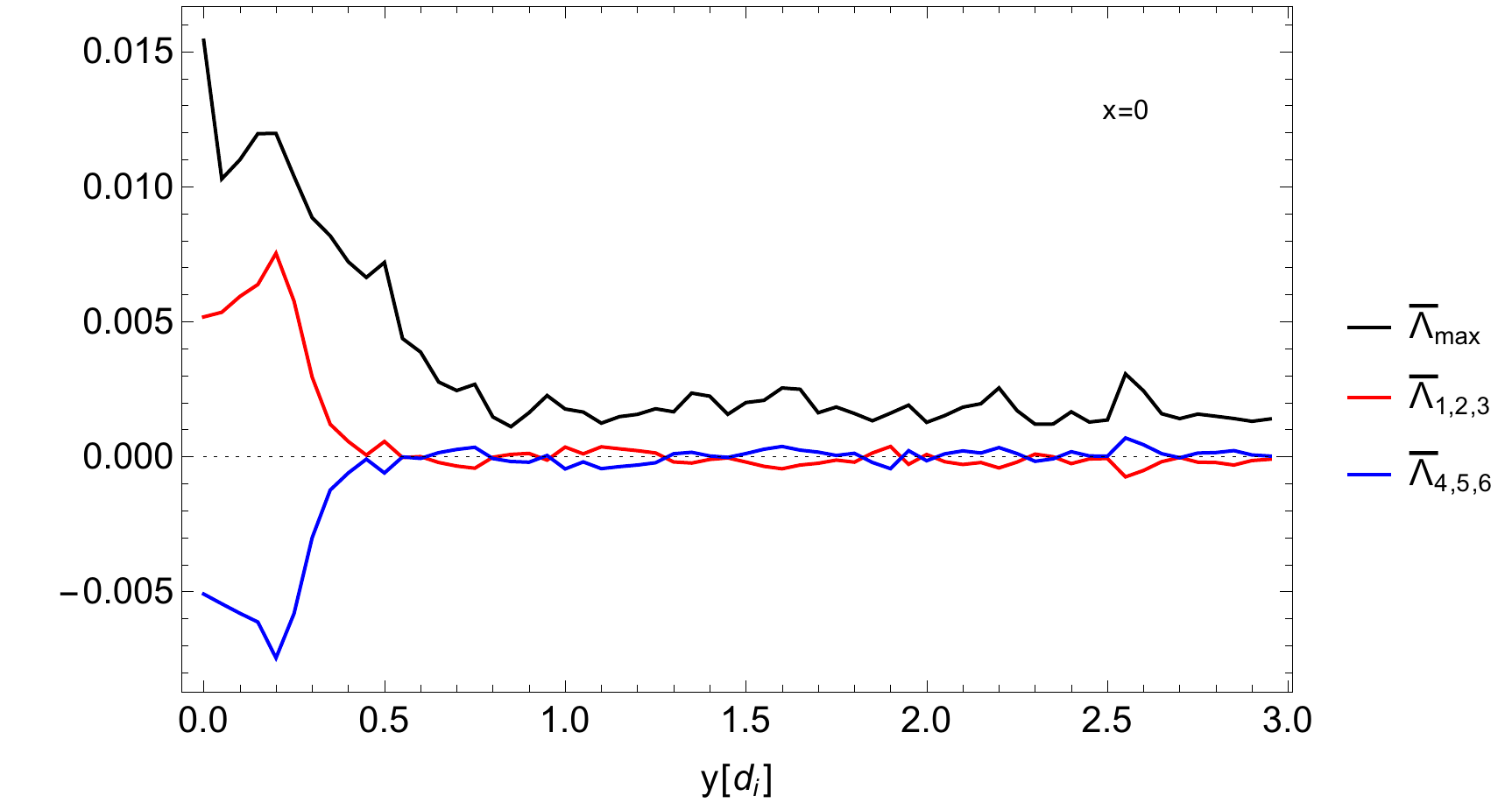}
\caption{$\bar\Lambda_{max}$, $\bar\Lambda_{123}$, and $\bar\Lambda_{456}$ for $y=0$ (left), 
and $x=0$ (right), respectively.}
\label{f:comp}
\end{center}
\end{figure}

In Fig.~\ref{f:LEN-LENv} we compare the average over $\bar\lambda_i$ with
$i=1,2,3$ ($\bar\Lambda_{123}$) with the average over $\bar\lambda_i$ with $i=4,5,6$
($\bar\Lambda_{456}$)
provided in color scale in the space $I_x\times I_y$. Here, the index $i$ refers
to the initial deviation vector $Y_i$, with $i=1,2,\dots,6$. Negative values are shown
in blue and correspond to $\bar\Lambda_{123}$ ($\bar\Lambda_{456}$) smaller than
unity while the quantities larger than unity are shown in red.  We clearly see
the complementary character of the two quantities: each region in red (blue) in
the upper figure has its correspondence in blue (red) in the bottom one. The symmetry
is due to the coupling of the generalized variables $\vec X(0)$ and the symmetric 
choice of the initial deviation vectors $\vec Y(0)$. A comparison of $\bar\Lambda_{max}$,
$\bar\Lambda_{123}$, and $\bar\Lambda_{456}$ is shown in Fig.~\ref{f:comp}. Notice the 
antisymmetry between $\bar\Lambda_{123}$ and $\bar\Lambda_{456}$ due to the symplectic 
structure of the dynamical problem.

\begin{figure}
\begin{center}
\includegraphics[width=.48\linewidth]{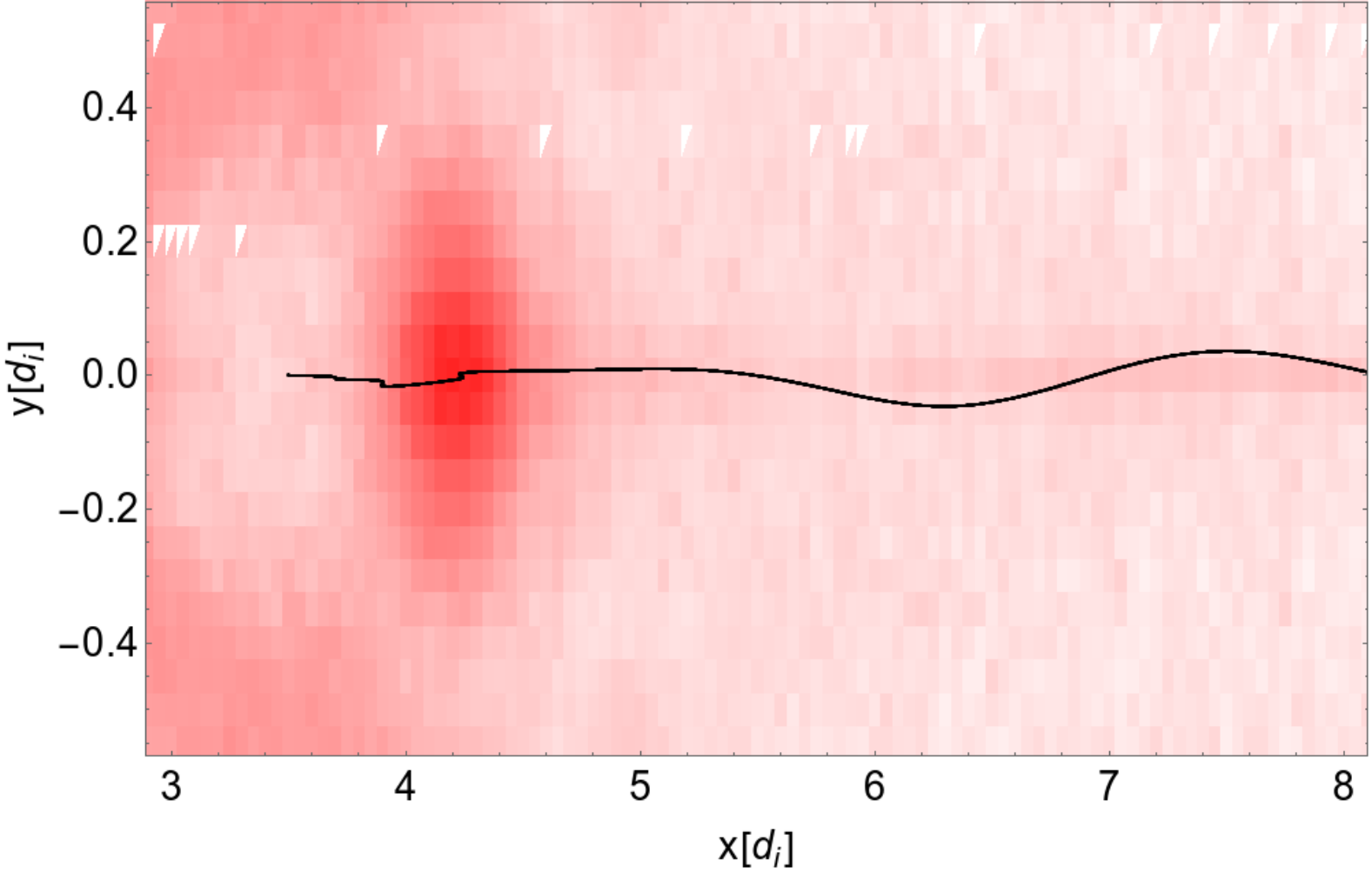}
\includegraphics[width=.48\linewidth]{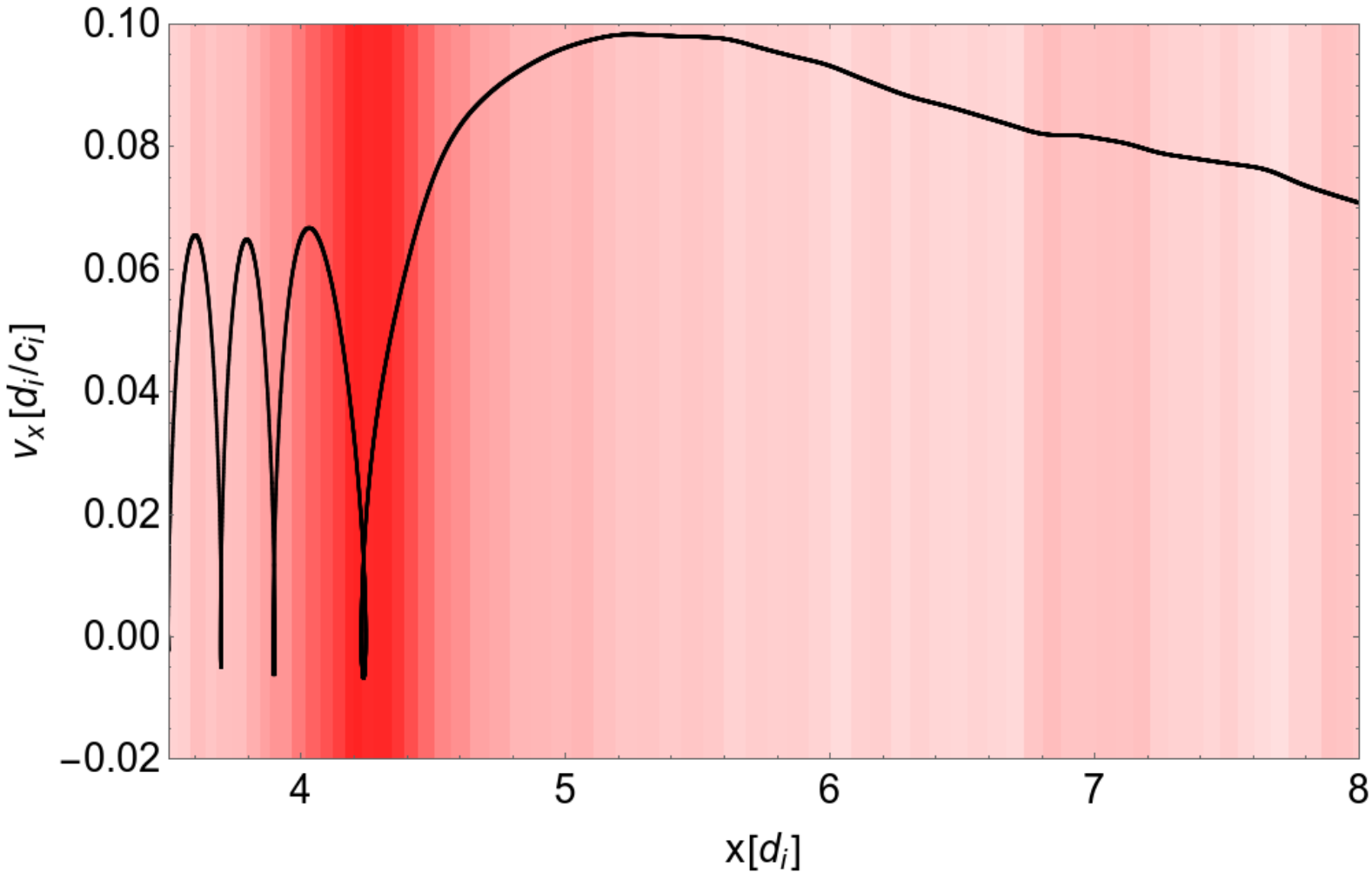} \\
\caption{Particle orbit (black) starting with zero velocity ad $y=0d_i$ at $x(0)=3.5d_i$.
Left: projection of the orbit to the $(x,y)$-plane; right: projection to $(x,v_x)$
(background shows maximum {\bf LEA}, the color code is the same as in 
Fig.~\ref{f:LEN}).}
\label{f:CLS}
\end{center}
\end{figure}

\begin{figure}
\begin{center}
\includegraphics[width=.48\linewidth]{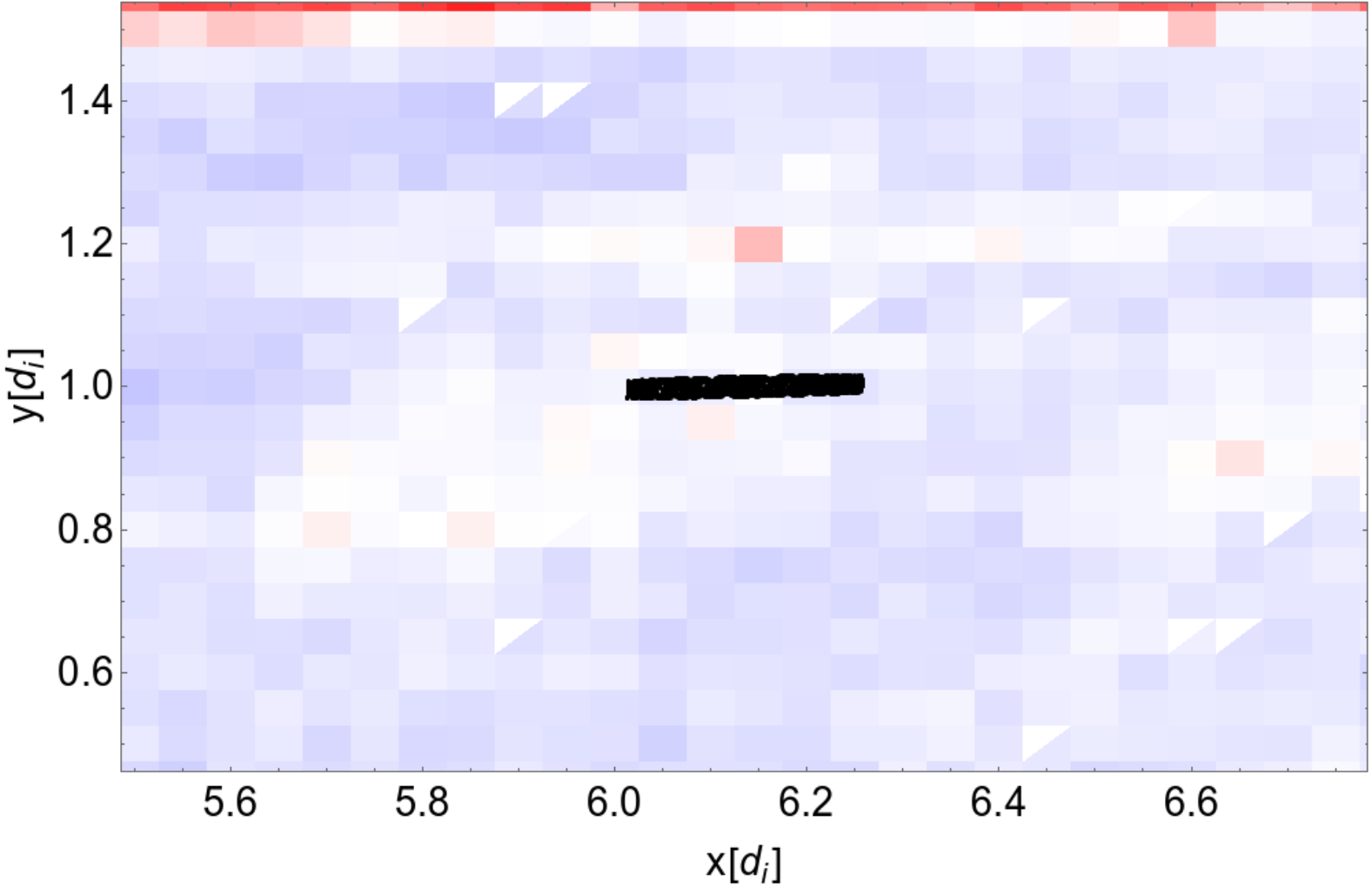}
\includegraphics[width=.48\linewidth]{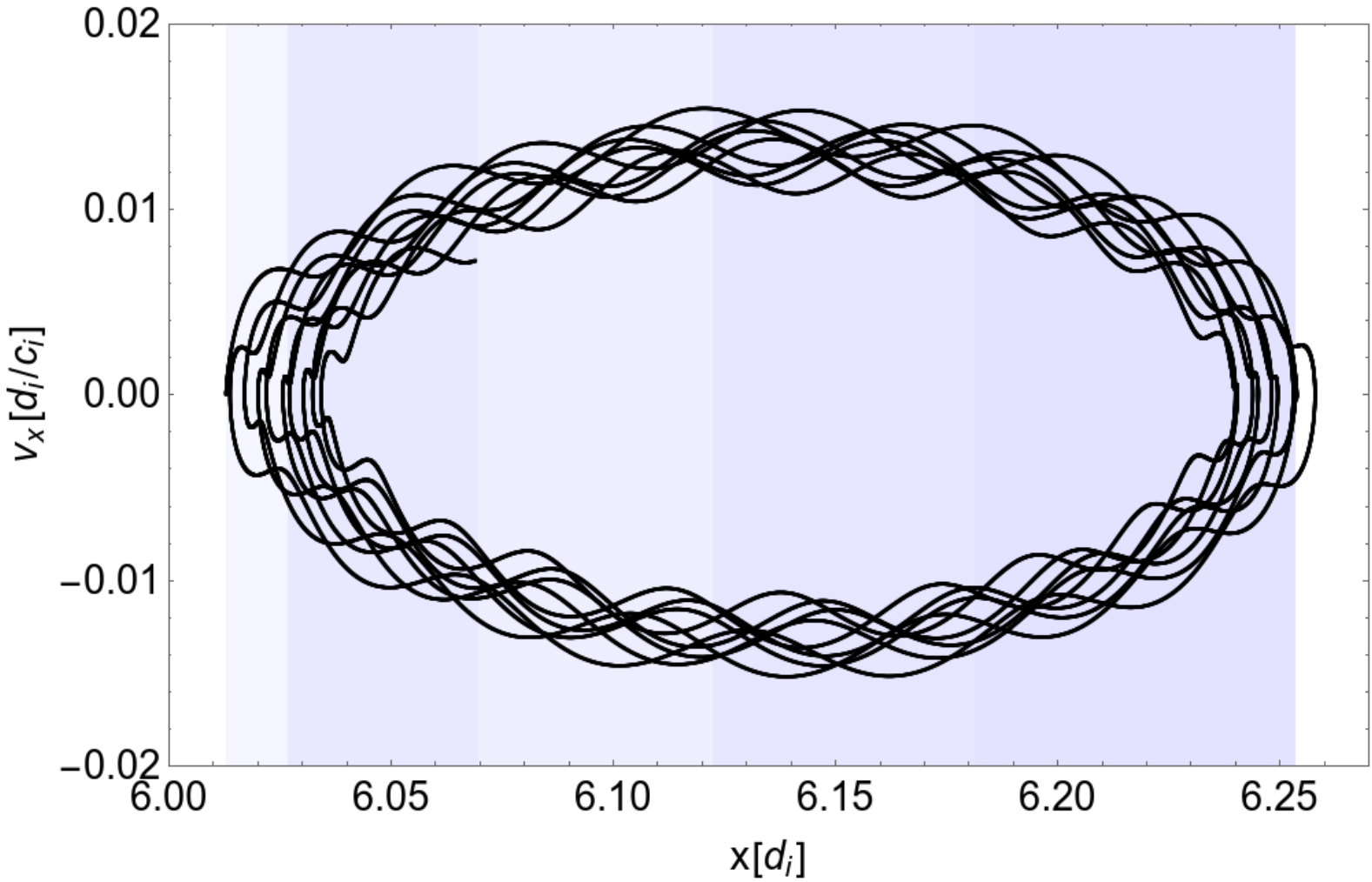} \\
\includegraphics[width=.48\linewidth]{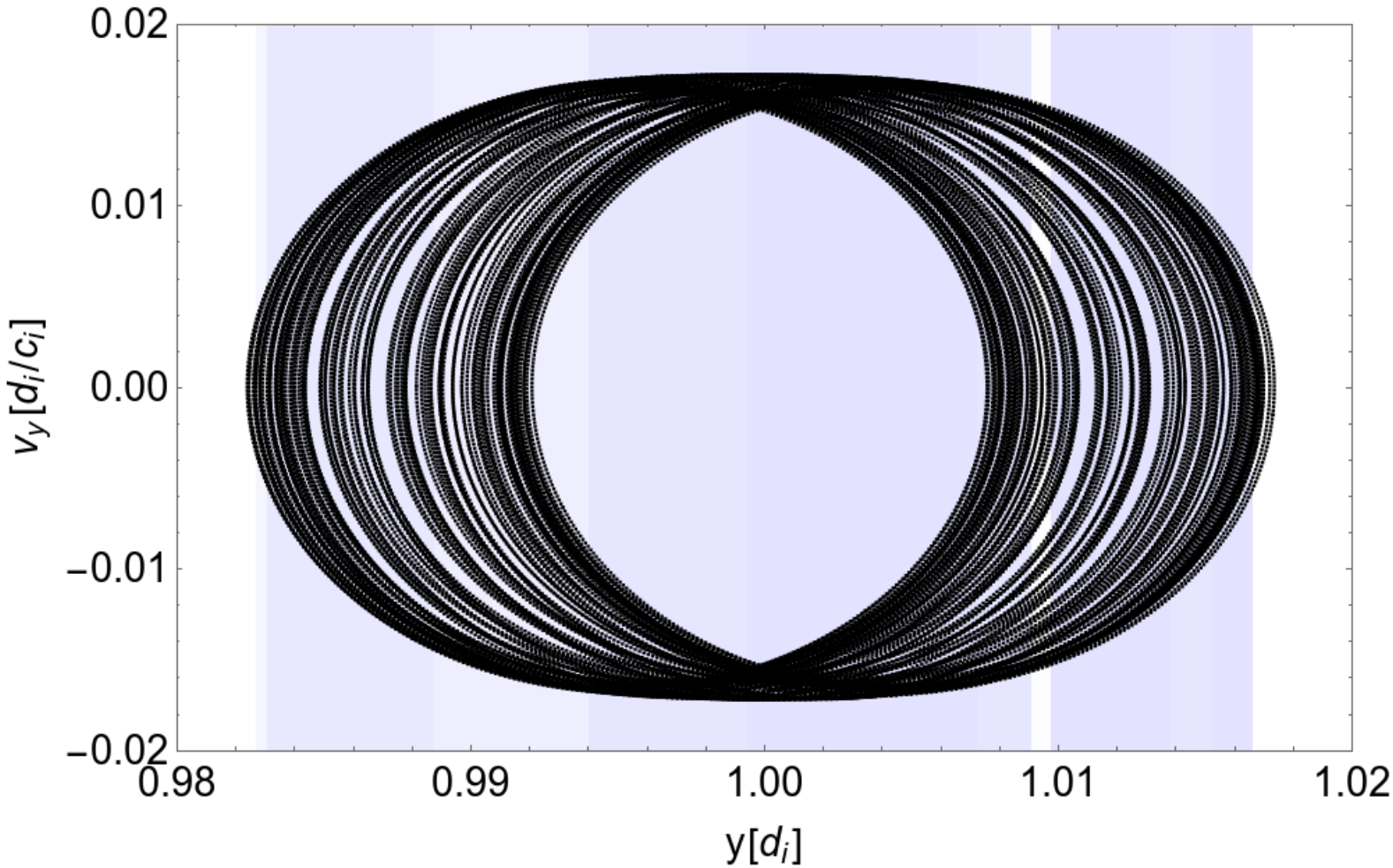}
\includegraphics[width=.48\linewidth]{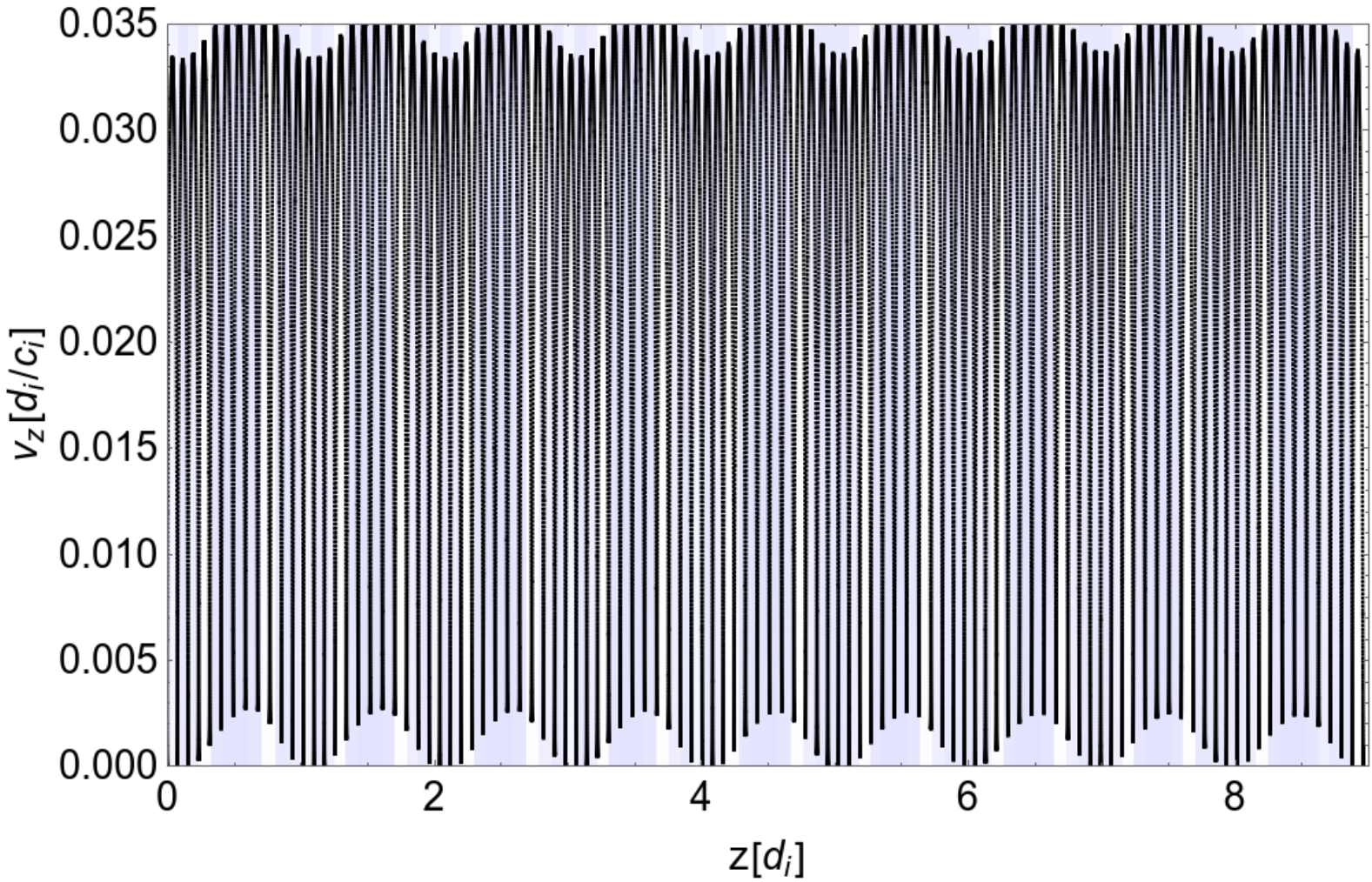} \\
\caption{Particle orbit (black) starting with zero velocity at $x(0)\simeq6.013d_i$
and $y(0)\simeq0.983d_i$. Projection to the $(x,y)$ (upper left), $(x,v_x)$ (upper right),
$(y,v_y)$ (lower left), and $(z,v_z)$-planes, respectively.
The background shows average of {\bf LEA} over velocity directions, the color code is the same as 
in Fig.~\ref{f:LEN-LENv}.}
\label{f:CLS2}
\end{center}
\end{figure}

\begin{figure}
\begin{center}
\includegraphics[width=.48\linewidth]{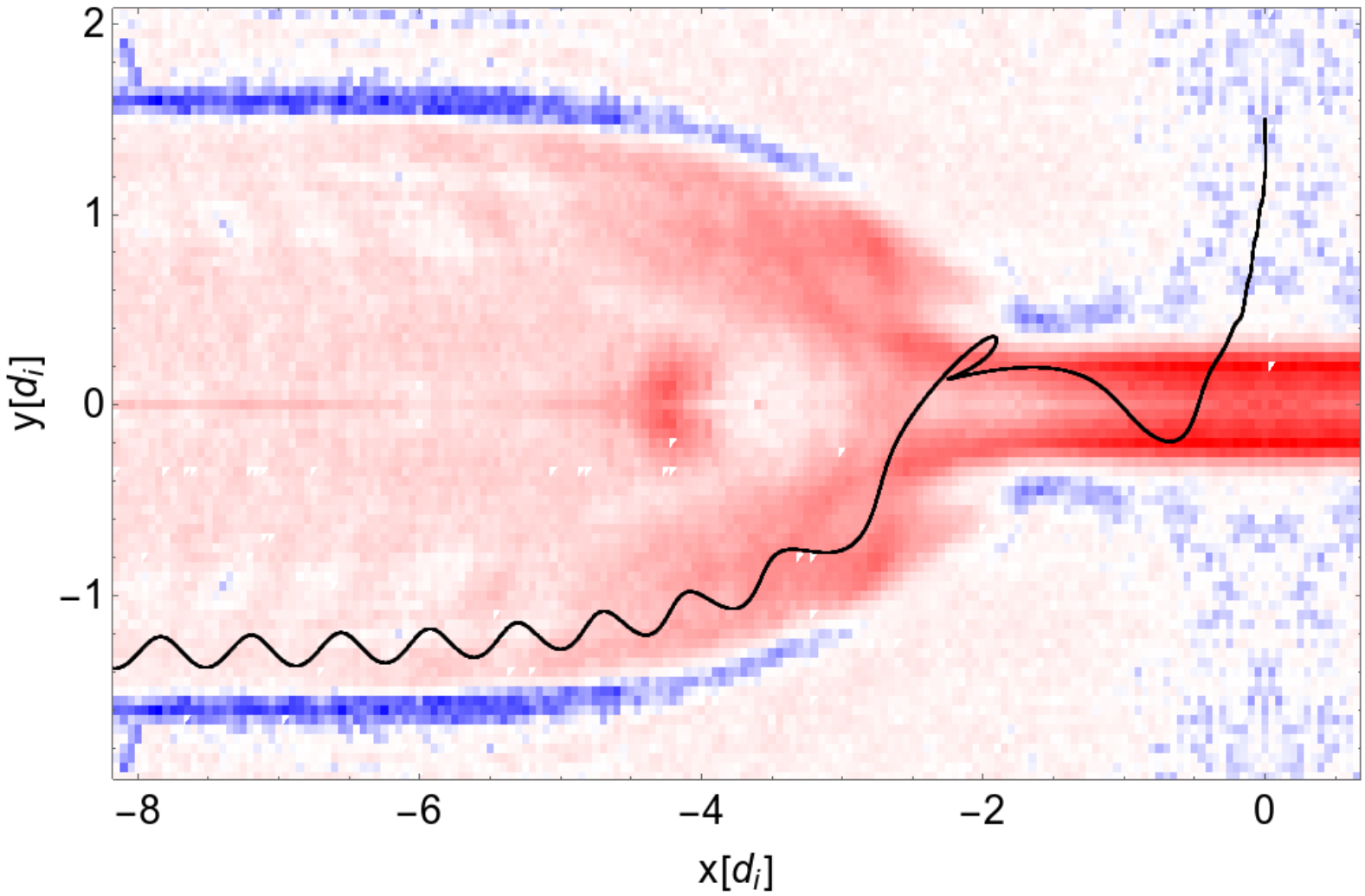}
\includegraphics[width=.48\linewidth]{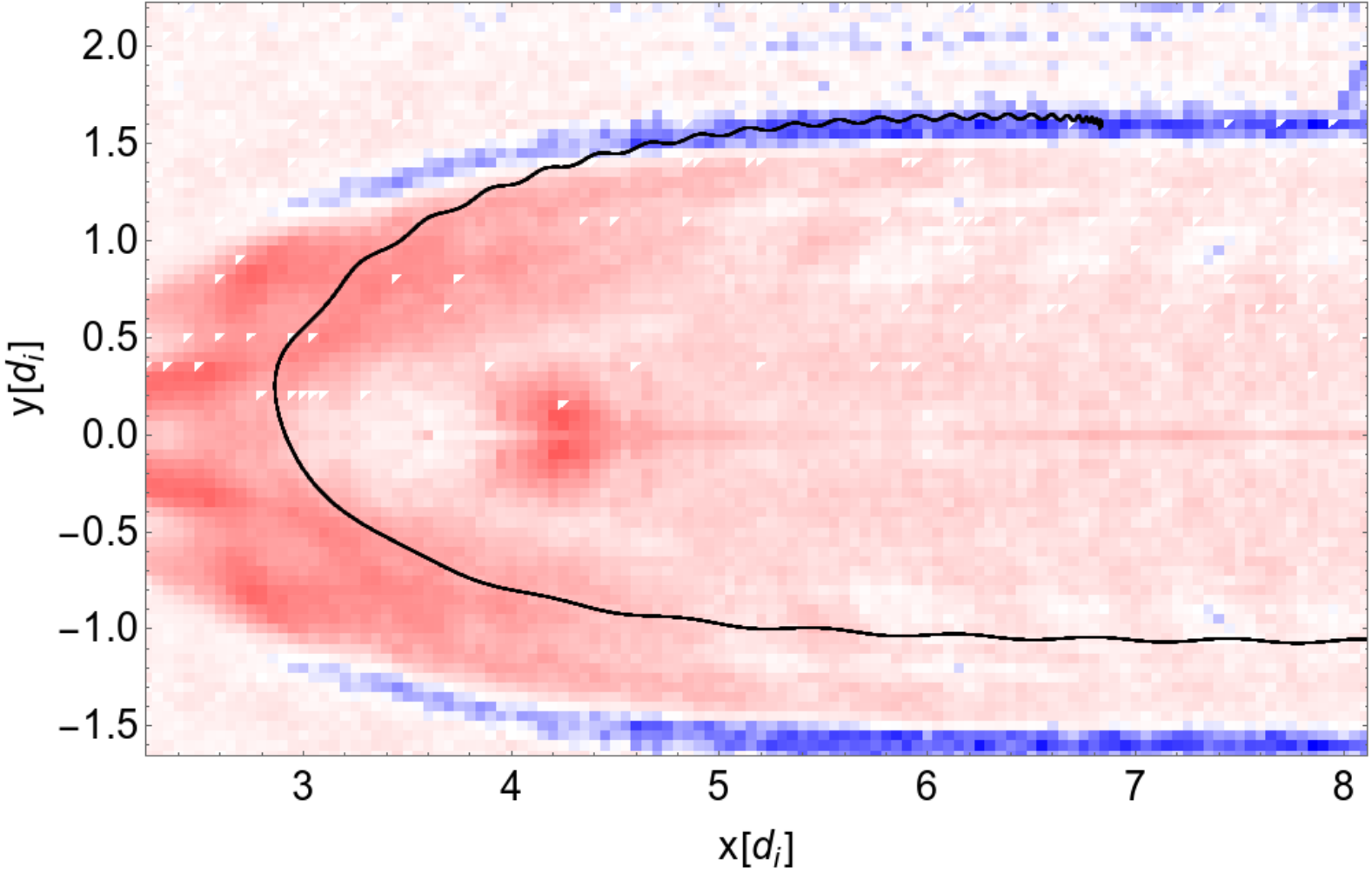} \\
\includegraphics[width=.48\linewidth]{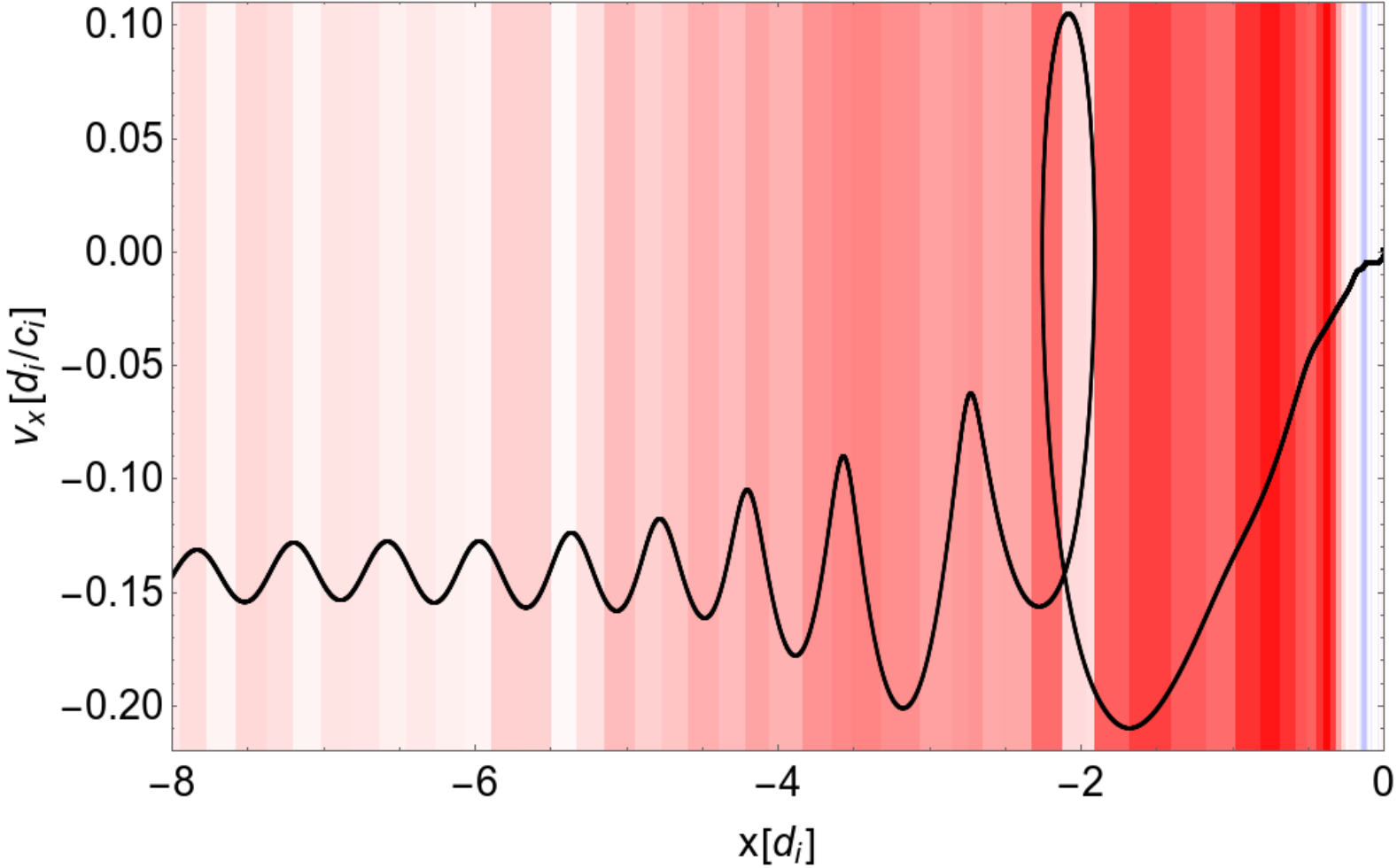}
\includegraphics[width=.48\linewidth]{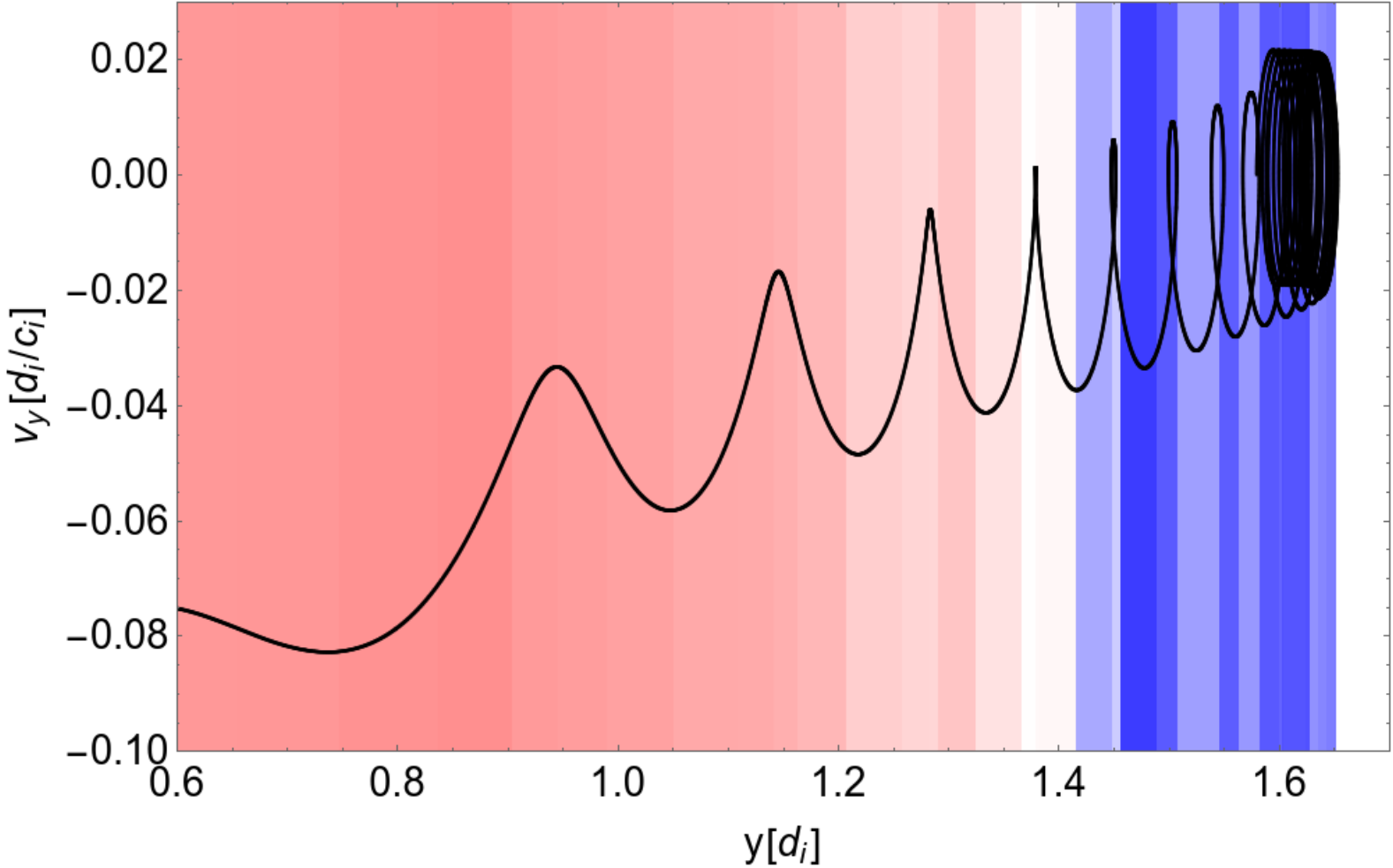} \\
\caption{Two distinct particle orbits that cross different regions of {\bf LEAs}.
Left: orbit starting at $x(0)=0d_i$, $y(0)=1.5d_i$ within the central inflow
region; projection to $(x,y)$-plane (top, left) and projection to the
$(x,v_x)$-plane (bottom, left). Right: orbit starting at $x(0)=6.826d_i$ and
$y(0)=1.580d_i$ at the border of the separatrix; projection to the $(x,y)$
(top,right) and $(y,v_y)$ (bottom, right) planes, respectively. Background
shows average of {\bf LEA} over position space the color code is the same as in
Fig.~\ref{f:LEN-LENv}.}
\label{f:CLS3}
\end{center}
\end{figure}

In classical theory the LCE and derived quantities are mainly used to
distinguish between regular and chaotic motions of individual particles in
phase space. In the present study we rather intend to use the {\bf LEA} to
characterize the mean influence of local perturbations on ensembles of particle
orbits at different locations close to the reconnection center. To demonstrate
the usefulness of the approach we show typical particle trajectories in
Fig~\ref{f:CLS}--\ref{f:CLS3}. In these figures, specific orbits are projected
to the $(x,y)$-plane, and as well as onto specific phase planes $(x,v_x)$,
$(y,v_y)$, and $(z,v_z)$, respectively. The color codes in the projections are
the same as in Figs.~\ref{f:LEN}-\ref{f:LEN-LENv}. In projections to phase planes
(e.g. $x$-$v_x$ in Fig.~\ref{f:CLS} and $y$-$v_y$, $z$-$v_z$ in Fig.~\ref{f:CLS2})
the color code indicates the local values of the {\bf LEA} at given position of the
particle orbit in the space $(x,y)$. As an example, the dark spot in the left panel
of Fig.~\ref{f:CLS} corresponds to the dark stripe in the right panel of the
same figure. Test particles are started with zero velocity at given location in the
$(x,y)$-plane. A case with large positive value of the maximum {\bf LEA} is shown in
Fig.~\ref{f:CLS}. The projection of the orbit onto the $(x,y)$-plane shows that
the orbit crosses a dark red spot located at $x\simeq4.2d_i$ and $y\simeq0d_i$ just
after its initial release at $x(0)=3.5d_i$. The effect of the crossing on the
shape of the orbit in phase space becomes clear when looking on the projection
of the orbit on the $(x,v_x)$-plane.  While in the light red region on the left
of the spot / stripe the drift along the $x$-direction is small, and regular kind of
oscillations along the velocity direction takes place, accelerations along the
$x$-direction strongly increases when the particle enters the dark red spot.
The effect completely destroys the oscillatory behaviour in the
$(x,v_x)$-plane, and drift in the $x$ direction becomes dominant. After exiting
the spot / stripe the effect of the perturbation becomes less efficient, and
the particle looses its velocity component along the $v_x$ direction,
before exiting the simulation box.

An example of the effect of negative values of the {\bf LEA} on particle orbits is
shown in Fig.~\ref{f:CLS2}. The orbit stays in the vicinity of its initial
release, but with increasing values in $z$. Motion takes place on
quasi-periodic curves in the $(x,v_x)$- and $(y,v_y)$-planes. As a
result, the orbit is trapped in rectangular region in its projection to the
$(x,y)$-plane, a phenomenon which is know to exist in plasma physics and called
'magnetic bottle'.

We notice that orbits shown in Fig.~\ref{f:CLS} and Fig.~\ref{f:CLS2} have carefully been chosen
to explain the effect of positive/negative values of the {\bf LEA} on the 
evolution of particle orbits. In our examples, the initial velocity and
out-of-plane direction have been set to zero for demonstration purpose only. 
It is clear that different orbits entering red or blue labeled cells 
close to the reconnection center will behave differently, i.e. experience different kinds of
accelerations depending on their actual dynamical state when crossing the regions. Typical examples
of such orbits are shown in Fig.~\ref{f:CLS3}. In both cases, the particle orbits enter regions 
with {\bf LEAs} of opposite sign. The darker the red regions and the stronger the magnitudes of the {\bf LEAs}, 
the stronger is the accelerations along the $x$ ($y$) directions. In regions with smaller magnitudes,
the drift in $x$ ($y$) directions is reduced, and oscillatory behaviour along the $v_x$ ($v_y$) 
planes takes place. The role of dark blue regions (related to negative values of {\bf LEAs}) is to
reduce drift along the $x$ ($y$) directions and to support oscillatory kind of behaviour of the 
orbit. As a direct consequence orbits remain longer in regions of magnetic reconnection which can 
be associated to negative values of the {\bf LEA}. We remark that we started this study using 
classical methods, e.g. the classical maximum Lyapunov Exponent method (mLE). The results were
unsatisfactory, and difficult to interpret for the following reason: while
some classical chaos indicators showed stable kinds of motion, individual particle trajectories
did not stay in the vicinity of its initial condition. Also, particle trajectories
were confined to regions in phase space for long time while the chaos indicator
predicted them to be unstable. This phenomenon can easily be explained when looking at the lower 
right of Fig.~\ref{f:CLS3}: the orbit shown in black stays the majority of its integration time 
close to its initial condition and performing quasi-periodic kinds of oscillations. Eventually,
the orbit escapes to the left and exits the simulation box already after very short time.
A classical chaos indicator would render the orbit to be unstable if one would
calculate, e.g. the mLE on the basis of the whole orbit. But, the same indicator would
render the orbit to be stable if one would exclude the last part of the orbit. The solution 
to the problem is the {\bf LEA}: the color code in Fig.~\ref{f:CLS3} clearly demonstrates that the 
single orbit behaves regular within the blue region, while it behaves chaotic in the red region.

\section{Summary \& Discussions}
\label{s:sumu}

Particle orbits in the vicinity of regimes of magnetic reconnection are subject
to concurrent magnetic and electric perturbations of different origins. It is a
natural step to investigate the effect of these kinds of perturbations on
particle orbits using tools of nonlinear dynamics. In classical nonlinear
problems, Lyapunov Characteristic Exponents have shown to be a useful tool to distinguish
between regular and chaotic kinds of motion in phase space. The exponents are based on
the concept of tangent space, i.e. on the variational equations of the
dynamical system in mind. In this work we study the dynamical problem of
particle orbits in regimes of magnetic reconnection in Earth's magneto-tail.
We are interested in the effect of different magnetic topologies on the change
of the geometry of particle  orbits in different regions of magnetic
reconnection.  For this reason we derive and make use of the system of
variational equations that govern the evolution of particle orbits in phase
space. Our particle simulations are based on electric and magnetic fields that
have been obtained using PIC simulations.  Next, we generalize the
concept of LEs to ensembles of particles, and introduce the Lyapunov Ensemble
{\bf Averages}. This quantity serves as a useful tool to distinguish between
regions of fast and slow accelerations along different directions in phase space,
in particular to locate regions of oscillatory orbits and 'quasi'-regular
kinds of motion. It is possible to define regions in phase
and configuration space, where ensembles of particle orbits are trapped for much
longer times with respect to regions indicating strong accelerations.

Our study is a first in a series of studies related to the problem of magnetic reconnection
from a dynamical systems point of view. In our ongoing study we investigate different
magnetic field configurations by the means of {\bf LEA} methods, and the possibility to derive 
additional macroscopic observables on the basis of the {\bf LEAs}. Our results may also serve
for a better interpretation of observations from space measurements in Earth's magneto-tail, e.g.
the Cluster, Themis, MMS and other mission.





\section*{Acknowledgements}

This work took benefit from the FWF project P30542-N27 (CL), and FWF project FWF S11608-N16 (EPL). 
We gratefully thank the plasma group at the Space Research Institute for fruitful discussions, i.e. 
R. Nakamura, T. Nakamura, Y. Narita, Y.L. Sasunov,  and Z. V\"or\"os. We thank an anonymous
reviewer for critical comments during the peer-reviewing process.


\bibliography{biblio}

\bibliographystyle{apj}

\end{document}